\documentclass[prb,aps,showpacs,citeautoscript, superscriptaddress,amsmath,amssymb,floatfix,twocolumn,dvipsnames]{revtex4-1}
\usepackage{graphicx}
\usepackage{xcolor}
\usepackage{ytableau}
\usepackage[colorlinks,bookmarks=false,citecolor=blue,linkcolor=red,urlcolor=black]{hyperref}
\usepackage{amsfonts}
\usepackage{amsmath}
\usepackage{times}
\usepackage{amssymb}
\usepackage{changes}
\usepackage{blkarray}
\usepackage{relsize}
\usepackage{bm}
\hypersetup{
    colorlinks=true,
    linkcolor=blue,
    citecolor=blue,    
    urlcolor=blue,
}

\begin{document}

\title{The SU(N) Fermi-Hubbard Model on two sites: Bethe Ansatz solution and Quantum Phase Transition of the Lipkin-Meshkov-Glick Model in the large-N limit.}
\author{Pierre Nataf}
\affiliation{Laboratoire de Physique et Mod\'elisation des Milieux Condens\'es, Universit\'e Grenoble Alpes and CNRS, 25 avenue des Martyrs, 38042 Grenoble, France}

\date{\today}
\begin{abstract} 
We show that the $\mathrm{SU}(N)$ Fermi-Hubbard model (FHM) on two sites, where $N$ is the number of flavors of each fermion, corresponds to an exactly solvable two-level many-boson model that R.W. Richardson analytically solved long time ago. We express the Bethe ansatz solutions as a function of the physical parameters of the  $\mathrm{SU}(N)$ FHM, and recast its eigenvalues and eigenstates in terms of the Richardson pair energies and creation operators.
In this context, the connection with the well studied Lipkin-Meshkov-Glick (LMG) model, known as equivalent to the Richardson model, is established and serves as a guideline to the prediction of some $N$-body physics phenomena in the two-sites  $\mathrm{SU}(N)$ FHM with $N$ particles. In particular, the LMG second order Quantum Phase Transition (QPT) is shown to occur in the $\mathrm{SU}(N)$ FHM for an attractive density-density interaction $U$ equal to $U_c=-1/(2N)$, in units of the (absolute value of the) tunneling amplitude between the two sites.
We show the finite size energies, the gap, and the kinetic energy which all reveal the transition, as a function of $U$ for values of $N$ from $N=3$ to $N=36$,
suggesting that the QPT could be experimentally achieved with current technologies involving $SU(N)$ ultracold atoms or molecules.
Finally, we show the entanglement entropy of the first site with respect to the second, and it scales like $N$ at the transition, in contrast with several two-modes models.

\end{abstract}

\maketitle


{\it Introduction.}
The Fermi-Hubbard model (FHM) is a paradigmatic model for interacting fermions on a lattice \cite{Hubbard_1963,Gutzwiller_1963}, which aids in understanding certain electronic and magnetic properties of strongly correlated electrons in materials \cite{Scalapino_2012}. 
In particular, the $\mathrm{SU}(2)$ FHM on the square lattice, which is believed to explain the pairing mechanism in cuprates superconductors \cite{Anderson_1987, Rice_1988,Scalapino_2012} has motivated plethora of theoretical investigations \cite{review_Arovas_2022,review_Corboz_2022}. 

 A generalization of the $\mathrm{SU}(N=2)$ FHM is the $\mathrm{SU}(N)$ FHM \cite{Assaraf_1999,honerkamp_ultrcold_2004,capponi_phases_2016,Ibarra_Garcia_Padilla_2024}, where $N$ is the number of degenerate orbitals. Through the large-N limit \cite{affleck_exact_1986,Affleck_1988,Rokhsar_1990,Marder_1990}, it was first used to give some asymptotic theoretical description of fermions with spins $1/2$.
 
 For finite $N>2$, it can also describe some condensed matter systems like transition metal compounds \cite{Khomskii_1998,Yamada_2018}, twisted bilayer \cite{zhang2021},
 or ultra-cold atoms trapped on various engineered optical lattices~\cite{wu_exact_2003,Wu_review_2006,gorshkov_two_2010,Cazalilla_2014}.
 In this latter case, the $\mathrm{SU}(N)$ symmetry is quasi-exact and results from the perfect decoupling of the electronic degrees of freedom from the nuclear spin \cite{gorshkov_two_2010,capponi_phases_2016,Ibarra_Garcia_Padilla_2024}. It can be tuned and be as large as $N=6$ for $^{173}$Yb and $N=10$ for $^{87}$Sr.
 The progress in the manipulation of $\mathrm{SU}(N)$  alkaline-earth atoms trapped on optical lattices are continuous \cite{taie_realization_2010,desalvo_degenerate_2010,zhang_spectroscopic_2014,pagano_one_2014,taie_su6_2012,Becker_2021,Fallani_2022}, and the experimental achievements are frequently based on the realization of the SU(N) FHM \cite{hofrichter_direct_2016,taie2020observation,Pasqualetti_2024}.
 
 From a computational point of view, the study of these models is challenging as the dimension of the full Hilbert space is:
 \begin{align}
 \label{dimension}
 \text{Dimension}=2^{NL},
 \end{align}
  where $L$ is the number of sites of the lattice.
 From such a dimension, it might be fruitful to imagine two opposite (non trivial) limits: the first one, small $N$, large $L$ has been the subject of most of the theoretical studies on the FHM.
 For $N=2$, it actually corresponds to the profusion of works made for the standard FHM for electrons with spins $1/2$ \cite{review_Arovas_2022,review_Corboz_2022}, while for small $N>2$ the theoretical studies are mostly collected in the recent review \cite{Ibarra_Garcia_Padilla_2024}.
 The opposite limit, large (but finite) $N$ and small $L$, has, however, not been so much investigated.
 
 The first non-trivial small number of lattices $L$ we should think about is $L=2$, and it is the purpose of the present letter to explore such a model, i.e the $\mathrm{SU}(N)$ FHM on $L=2$ sites (FHM2S).
 As we show below, it admits a Bethe ansatz solution due to Richardson \cite{Richardson}, and shares some features of the $N$-body physics of the Lipkin-Meshkov-Glick (LMG) model\cite{LMG_1965}, as we can map the $\mathrm{SU}(N)$ FHM2S onto the LMG model. From such a mapping, we will predict a second order Quantum Phase Transition (QPT) in the attractive region of the density-density interaction $U$, in the large $N$ limit.
Such a limit is all the more relevant from a physical point of view that it has recently been proposed to realize the $\mathrm{SU}(N)$ FHM with {\it shielded} ultracold molecules, for which 
$N$ can go up to $N=36$ with Na$^{40}$K \cite{Mukherjee_2024}. 
 
 We will also calculate the entanglement entropy of the first site with respect to the second, as this quantity is often taken as an indicator of a QPT\cite{Osterloh_2002,Osborne_2002,Vidal_2003,Filippone_2011}, before
giving some experimental considerations and perspectives of our work.

{\it The $\mathrm{SU}(N)$ FHM on two sites.}
The Hamiltonian of the $\mathrm{SU}(N)$ FHM2S reads:
\begin{equation}
\label{Hamiltonian}
H= -t \Big{(}  E_{12}+ E_{21}\Big{)} + \frac{U}{2} \Big{(} E_{11}^2+E_{22}^2 \Big{)}, 
\end{equation}
where $t$ is the hopping amplitude between site $1$ and $2$, and the density-density on-site interaction amplitude is $U$. 
The $\mathrm{SU}(N)$ invariant hopping terms are ($ 1 \leq i,j \leq 2$):
\begin{equation}
\label{Hopping}
E_{ij}=E_{ji}^{\dag}=\sum_{\sigma=1}^N c^{\dag}_{i \sigma}c_{j \sigma},
\end{equation}
where $c^{\dag}_{i \sigma}$ (resp. $c_{i \sigma}$ ) is the creation (resp. annihilation) operator of a fermion of color $\sigma$ on site $i$ ($i=1,2$). 
These operators  satisfy the commutation relation of the generators of the $U(2)$ algebra ($\forall {1 \leq i,j,k,l \leq 2}$) :
\begin{equation}
\label{commutation}
[E_{ij},E_{kl}]=\delta_{jk}E_{il}-\delta_{li}E_{kj}.
\end{equation}
From this relation, one shows that the total number of fermions, i.e $E_{11}+E_{22}$, commutes with $H$, and we call $M$ the conserved number of fermions, i.e $E_{11}+E_{22}\equiv M$.

Being $SU(N)$ symmetric, the Hamiltonian $H$ also leaves invariant the sectors of the full Hilbert space that are labelled by the {\it Young Diagrams} (YD) or {\it shapes} $\boldmath{\alpha} = [\alpha_1,\alpha_2,...,\alpha_{M_1}]$, with $M_1$ the number of rows of the diagram ($1 \leq M_1 \leq N$) such that $\alpha_1\geq \alpha_2 \geq...\geq \alpha_{M_1}\geq 1 $, and which represent the irreducible representations (irreps) of $SU(N)$. 

\begin{figure}
\centerline{\includegraphics[width=.46\textwidth]{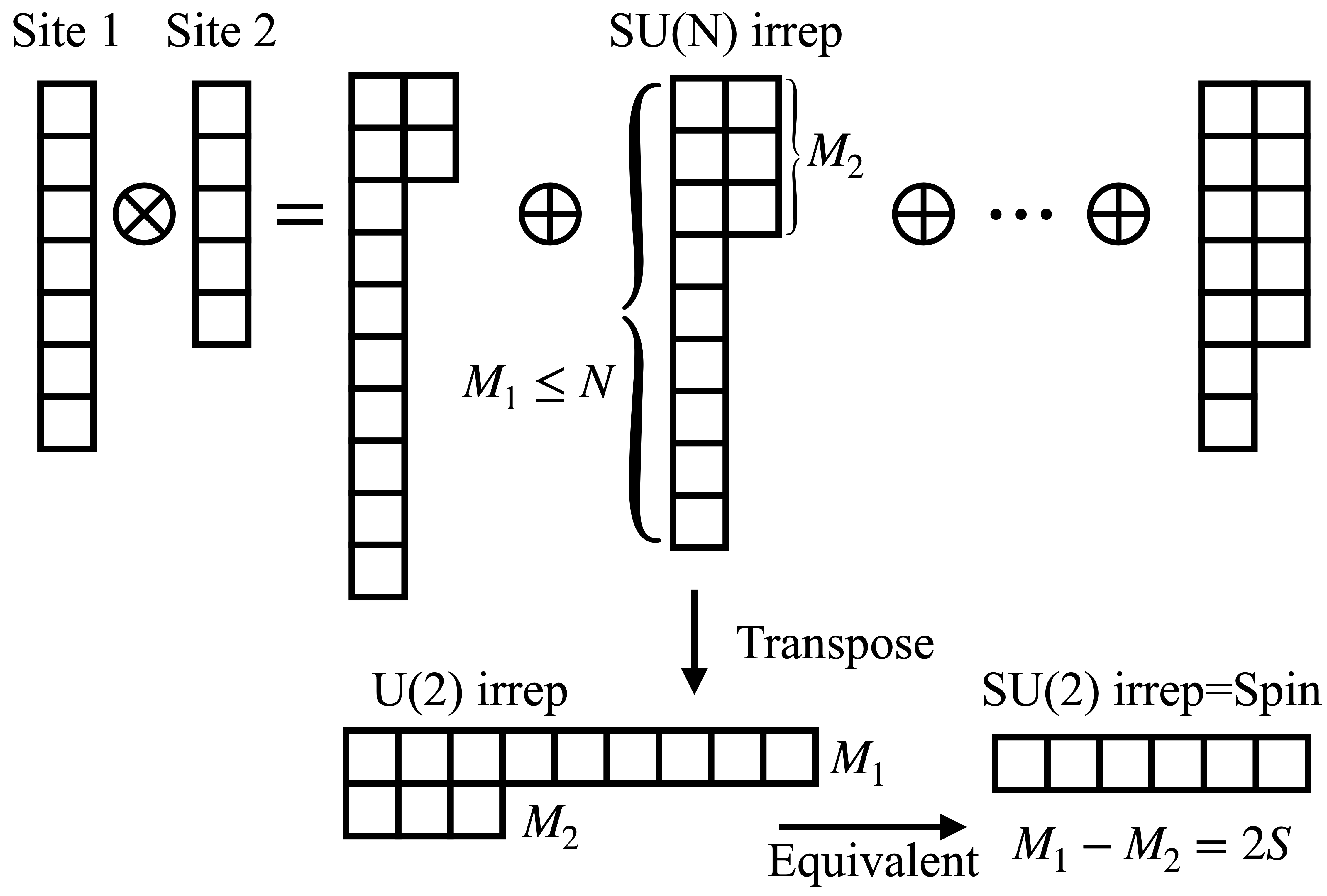}}
\caption{Example of relevant Young Diagrams (YD) for the SU(N) Fermi-Hubbard model on two sites (FHM2S).
 A general YD should be $\alpha = \alpha^{M_1}_{M_2}=[2,\cdots , \alpha_{M_2}=2, \alpha_{M_2+1}=1, \cdots  \alpha_{M_1}=1]$, where $M_2$ is the number of 2-boxes rows and $M_1\leq N$ is the total number of rows.
 By transposing $\alpha = \alpha^{M_1}_{M_2}$, one obtains $\bar{\alpha} = \bar{\alpha}^{M_1}_{M_2}$ which represents the $U(2)$ irrep on which one should calculate the generators $E_{ij}$ to compute the matrix of the Hamiltonian on the $SU(N)$ sector $\alpha$ \cite{Botzung_2023_PRL}. For two sites, the transposed irreps are spins of length $S=(M_1-M_2)/2$ and the $SU(N)$ FHM2S Hamiltonian reduces to the spin Lipkin-Meshkov-Glick (LMG) Hamiltonian.
}
\label{fig_irrep}
\end{figure}

In this framework, the number of particles $M=E_{11}+E_{22}$ is equal to the number of boxes, i.e. $\sum_{i=1}^{M_1} \alpha_i=M$.
For our current two-sites model, the relevant shapes $\alpha$ have at most two columns and $N$ rows (i.e $\alpha_1\leq 2$ and $M_1 \leq N$), as it can either be seen from the general protocole established in \cite{Botzung_2023_PRL,Botzung_2023}, either be understood from the tensor product\cite{itzykson_unitarity_1966} of two one-column irreps (with at most $N$ boxes in each), which individually represent the fully antisymmetric fermionic wave-functions on each site (cf Fig. \ref{fig_irrep}).

Let's define $\mathcal{H}^{M, N}_2$, the Hilbert space  for $M$ SU($N$) fermions on two sites interacting through the Hamiltonian in Eq.~\eqref{Hamiltonian}. 
It was shown in \cite{Botzung_2023_PRL,Botzung_2023} that $\mathcal{H}^{M, N}_2$ can be decomposed
as
\begin{equation}
\label{eq: decomposition_hilbert}
\mathcal{H}^{M, N}_2=\underset{\alpha}{\oplus}\overset{d^{\alpha}_N}{\underset{k=1}{\oplus}} \mathcal{H}^{\bar{\alpha},k}_2,
\end{equation}
where the outer sum runs over all the M-boxes YD $\alpha$ of maximum $2$ columns and $N$ rows. 
For a given $\alpha$, there are $d^{\alpha}_N$ independent sectors $\mathcal{H}^{\bar{\alpha},k}_2$ (for $k=1 \cdots d^{\alpha}_N$) which are invariant under the action of the Hamiltonian $H$. They are isomorphic with each other, with the same dimension $d^{\bar{\alpha}}_2$, 
where $\bar{\alpha}$ is the \textit{transpose} of a YD $\alpha$, transforming its rows into columns [cf. Fig.~\ref{fig_irrep}].
For  $1 \leq k \leq d^{\alpha}_N$, the sector $\mathcal{H}^{\bar{\alpha},k}_2$ realizes the irrep $\bar{\alpha}$ of the $U(2)$ algebra, the operators $E_{ij}$ are the generators of (for $ 1 \leq i,j \leq 2$).
Since $1 \leq k \leq d^{\alpha}_N$, they give rise to some multiple copies (some {\it multiplicities}) of the eigenspectrum corresponding to the irrep $\alpha$ in the full eigenspectrum of the Hamiltonian $H$ (cf Eq~\eqref{Hamiltonian}) on $\mathcal{H}^{M, N}_2$.
In addition, the dimension $D^{M,N}_2$ of the Hilbert space $\mathcal{H}^{M,N}_2$ is $
D^{M,N}_2\equiv \text{dim} (\mathcal{H}^{M, N}_2)=\sum_{\boldmath{\alpha}} d^{\alpha}_N  d^{\bar{\alpha}}_2$ and is such that
$\sum_{M=0}^{M=2N} D^{M,N}_2=2^{2N}$, as it should be when we sum up the dimensions $D^{M,N}_2$ for all the different possible number of particles $M$.
Moreover, the quantity $d^{\alpha}_N $ is the dimension of the $\mathrm{SU}(N)$ irrep $\alpha$  that one can obtain using, e.g., the hook length formulas~\cite{Weyl1925Dec, Robinson1961}.
For a two-columns shape $\alpha \equiv \alpha^{M_1}_{M_2}=[2,2,\cdots , \alpha_{M_2}=2, \alpha_{M_2+1}=1, 1, \cdots  \alpha_{M_1}=1]$ (where $M_2$ is the number of 2-boxes rows and $M_1$ is the total number of rows), a simple application of this formula leads to:
\begin{equation}
d^{\alpha^{M_1}_{M_2}}_N= \binom{N}{M_1}\binom{N+1}{M_2}  \frac{M_1-M_2+1}{M_1+1},
\end{equation}
where $\binom{p}{q}=p!/(q!(p-q)!)$ is the binomial coefficient. 
The integer $d^{\alpha^{M_1}_{M_2}}_N$ is the multiplicity of each eigenvalue coming from the diagonalization of the Hamiltonian $H$ on an individual sector $\mathcal{H}^{\bar{\alpha},k}_2$. 
In the Tab. \ref{table: table0}, we give the multiplicities $d^{\alpha^{M_1}_{M_2}}_N$ for $N=M=10$.

\begin{table}[h]
    \centering
\begin{tabular}{|c|c|c|c|c|c|c|} 
\hline
$(M_1,M_2)$&(10,0)&(9,1)&(8,2)&(7,3)&(6,4)&(5,5)\\
\hline

$d^{\alpha^{M_1}_{M_2}}_N$ & 1&99&1925&12375&29700&19404\\
\hline
\end{tabular}
\caption{Multiplicities $d^{\alpha^{M_1}_{M_2}}_N$ for $N=10$ flavors and $M=10$ particles.
The relevant two-columns $SU(N)$ irreps $\alpha^{M_1}_{M_2}$ are characterized by the two integers $(M_1 \geq M_2)$ such $M=M_1+M_2$,
which also appear in Fig. \ref{fig_Energies_SU10} under their pictorial form. Note that $(M_1,M_2)=(10,0)$ correspond to the $SU(10)$ singlets irrep.}
\label{table: table0}
\end{table}

In \cite{Botzung_2023_PRL,Botzung_2023}, we explained how to diagonalize $H$ on each sector $\alpha$, using  the basis of semi-standard Young Tableaux (SSYT) of shape $\bar{\alpha}$ and the Gelfand Tstelin representation of the generators $E_{ij}$.

We apply the general procedure of \cite{Botzung_2023_PRL,Botzung_2023} to the FHM2S Hamiltonian. Then, for a two-columns shape $\alpha^{M_1}_{M_2}$, the transposition gives $\bar{\alpha}^{M_1}_{M_2}=[M_1,M_2]$ which is a two-rows irrep of $U(2)$, which behaves as a spin of length $M_1-M_2$, which is a one-row irrep of $SU(2)$. In fact, calling $E_{ij}^{[M_1,M_2]}$ the matrices representing $E_{ij}$ on  the $U(2)$ irrep $[M_1,M_2]$ (for $1 \leq i,j \leq 2$), one has 
$E_{i,j}^{[M_1,M_2]}=E_{i,j}^{[M_1-M_2,0]}+ \delta_{i,j} M_2 \mathbb{I} $, where $\mathbb{I}$ is the identity matrix of dimension $M_1-M_2+1$.

Thus, for a given $\alpha=\alpha^{M_1}_{M_2}$, the SU(N)-symmetry resolved diagonalization of $H$ (cf Eq. \ref{Hamiltonian}) is equivalent to the diagonalization
of the Lipkin-Meshkov-Glick (LMG) spin Hamiltonian \cite{LMG_1965}:
\begin{equation}
\label{Hamiltonian_LMG}
H= -2t S_x + U S_z^2, 
\end{equation}
where we let aside the constant $M^2 U/4$, and where the operators $S_z \equiv (1/2)(E_{22}-E_{11})$, and $S_x =(1/2)(S_{+}+S_{-})$, with $S_{+}\equiv E_{21}=(E_{12}^{\dag})\equiv S_{-}^{\dag}$
are the standard ($M_1-M_2+1$-dimensional) operators for a spin of size $S \equiv (M_1-M_2)/2$ (e.g $M_1-M_2=2$ boxes correspond to spin $S=1$). 

In the following, we first review the connection between the LMG model and the Richardson pair bosonic Hamiltonian, in order to express the Bethe ansatz solutions in terms of the physical parameters of the SU(N) FHM2S (i.e. $M$, $U$ and $t$), and then we study some of its main properties, with in particular the occurence of a second order QPT, as a function of $U$ and $N$.

{\it Bethe Ansatz Solution.}
To match with Richardson's notations, we first perform a simple spin rotation \footnote{One can transform $H$ into $W^{\dagger}HW$, with $W=e^{i\frac{\pi}{2}S_x}e^{-i\frac{\pi}{2}S_y}$} to rewrite $H$ in Eq. \ref{Hamiltonian_LMG} as 
\begin{equation}
\label{Hamiltonian_LMG_2}
H= -2t S_z +  U S_y^2
\end{equation}
 We then introduce the Jordan-Schwinger bosonic representation of the spin operators:
\begin{align} \label{Jordan_Schwinger}
&S_{+}=a^{\dagger}_{\uparrow}a_{\downarrow}, \hspace{.5cm}S_{-}=a^{\dagger}_{\downarrow}a_{\uparrow}, \hspace{.5cm}S_z=\frac{1}{2}(n_{\uparrow}-n_{\downarrow}), \\
&\text{with} \hspace{1cm} n_{\uparrow}=a^{\dagger}_{\uparrow}a_{\uparrow}, \hspace{1.5cm} n_{\downarrow}=a^{\dagger}_{\downarrow}a_{\downarrow} \nonumber.
\end{align}
The creation and annihilation operators $a^{\dagger}_i$ and $a_j$ are independent bosonic operators satisfying $[a_{i},a^{\dagger}_j]=\delta_{i,j}$  and $[a_i,a_j]=[a^{\dagger}_i,a^{\dagger}_j]=0$ for $i,j=\uparrow,\downarrow$. Then, one can rewrite the Hamiltonian as:

\begin{align}
\label{Richardson_Hamiltonian}
H=t n_{\uparrow}-t n_{\downarrow}+\frac{-U}{4} \sum_{l,l'=\uparrow,\downarrow} (a^{\dagger}_{l})^2 a_{l'}^2 +\frac{U}{4}(n_{\uparrow}+ n_{\downarrow})^2 
\end{align}
The last term of the previous Hamiltonian is constant for states living in the one-row SU(2) irrep of length equal to $2S$, as the total number of bosons determines the SU(2) irrep, i.e $n_{\uparrow}+ n_{\downarrow} =2S$.
The sum of the first three terms of $H$ in Eq. \ref{Richardson_Hamiltonian}, that we call from now on $H_R$ (so that $H=H_R+\frac{U}{4}(n_{\uparrow}+ n_{\downarrow})^2$), constitues a particular "two level" case of the Richardson's bosonic pair Hamiltonian (cf. Eq. 2.2 in \cite{Richardson} where $\epsilon_{\uparrow}=-\epsilon_{\downarrow}=t$ and $g=-U/2$).
For self-consistency, we summarize the main steps of Richardson's derivation towards the Bethe equations, also detailed in \cite{Pan_1999}.

First, we introduce four states $\vert \varphi _{\bm{\nu}} \rangle$, where $\bm{\nu} \equiv (\nu_{\uparrow},\nu_{\downarrow})$, with $\nu_{\uparrow},\nu_{\downarrow}=0,1$,
and defined as:
\begin{align}
&\vert \varphi _{(0,0)} \rangle=\vert \vert 0 \rangle, \hspace{1cm}\vert \varphi _{(1,1)} \rangle=a^{\dagger}_{\uparrow}a^{\dagger}_{\downarrow}\vert \vert 0 \rangle, \label{lowest_states_1} \\
 &\vert \varphi _{(0,1)} \rangle=a^{\dagger}_{\downarrow}\vert \vert 0 \rangle,\hspace{.7cm}\vert \varphi _{(1,0)} \rangle=a^{\dagger}_{\uparrow}\vert \vert 0 \rangle,  \label{lowest_states_2}
\end{align}
where $\vert \vert 0 \rangle$ is the vacuum for the two bosonic modes.
These states nullify the $a_{l}^2$ operators: $a_{l}^2\vert \varphi _{\bm{\nu}} \rangle=0$ for ($l=\uparrow, \downarrow$) and are eigenstates of $H_R$ with eigenvalues
$E_{\bm{\nu}}= t(\nu_{\uparrow}-\nu_{\downarrow})$:
\begin{equation}
H_R \vert \varphi _{\bm{\nu}} \rangle=E_{\bm{\nu}}\vert \varphi _{\bm{\nu}} \rangle.
\end{equation}

Richardson's solution consists in looking for general (unnormalized) eigenstates $\vert \Psi \rangle $ of $H_R$ of the forms: 
\begin{equation}
\label{general_eigenstate}
\vert \Psi \rangle=B_1^{\dagger} \cdots B_p^{\dagger} \vert \varphi _{\bm{\nu}} \rangle,
\end{equation}
where (for $1 \leq q \leq p $):
\begin{equation}
\label{B_q}
B_q^{\dagger}= u_q^{\uparrow}(a^{\dagger}_{\uparrow})^2 +u_q^{\downarrow} (a^{\dagger}_{\downarrow})^2.
\end{equation}
The operators $B_q^{\dagger}$, in which the coefficients $u_q^{\uparrow,\downarrow}$ will be determined through the next Bethe ansatz Equation (BAE),
create a coherent superposition of pairs of particles, so that the general eigenstate $\Psi$ in Eq. \ref{general_eigenstate}, contains $p$ pairs of particles, and
$n_{p,\bm{\nu}}\equiv 2p+\nu_{\uparrow}+\nu_{\downarrow}$ particles in total.
For a given spin equal to $S$, one has $2S=n_{p,\bm{\nu}}$ particles and it determines both the number $p$ and the possible initial states $\vert \varphi _{\bm{\nu}} \rangle$ on which the $B_q^{\dagger}$  act as shown in Tab. \ref{table: table1}
\begin{table}[h]
    \centering
\begin{tabular}{|c|c|c|} 
\hline
$2S=M_1-M_2$&$ \bm{\nu}=(\nu_{\uparrow},\nu_{\downarrow})$&$p$ (number of pairs)\\
\hline
\text{odd} & (1,0) \hspace{0.05cm} \text{or}\hspace{0.1cm}(0,1)&$S-1/2$\\
\hline
\text{even} & (0,0) \hspace{0.05cm} \text{or}\hspace{0.1cm}(1,1)&$S\hspace{0.1cm} \text{or}\hspace{0.1cm}S-1$\\
\hline
\end{tabular}
 \caption{Quantum number $\bm{\nu}=(\nu_{\uparrow},\nu_{\downarrow})$ and number of pairs $p$ as a function of the spin size $S$.
}
\label{table: table1}
\end{table}

Thus, for a given spin $S$, the sector will be divided into two subspaces, according to the two possible  $\vert \varphi _{\bm{\nu}} \rangle$ that the bosonic pair creation operators act on,
as a reminiscence of the $\mathbb{Z}_2$ symmetry of the LMG model. \footnote{One has for instance $[H,W]=0$ with $W=e^{i\pi S_x}$ for $H$ defined in Eq. \ref{Hamiltonian_LMG}.}

The eigenvalue problem for the eigenstates shown in Eq. \ref{general_eigenstate}, take the form:
\begin{align}
&H_R \vert \Psi \rangle=E \vert \Psi \rangle \hspace{.4cm} \text{with} \hspace{.2cm} E=\sum_{q=1}^p E_q+t(\nu_{\uparrow}-\nu_{\downarrow}) \hspace{.2cm} \nonumber \\
& \text{if}  \hspace{.2cm} [H_R, B_1^{\dagger} \cdots B_p^{\dagger}] \vert \varphi _{\bm{\nu}} \rangle= (\sum_{q=1}^p E_q) B_1^{\dagger} \cdots B_p^{\dagger} \vert \varphi _{\bm{\nu}} \rangle. \label{eigenvalue_probem}
\end{align}
Working on the commutator appearing in Eq. \ref{eigenvalue_probem}, Richardson shown that the set of
 pair energies $\{E_q\}_{q=1 \cdots p}$ for a given integer $p$ and couple $\bm{\nu}= (\nu_{\uparrow},\nu_{\downarrow})$ should satisfy the system of coupled BAE ( $\forall \, 1 \leq q \leq p$):
 \begin{equation}
 \label{BAE}
 \sum_{1 \leq q' \neq q \leq p} \frac{4}{E_{q'}-E_q}+\frac{1+2\nu_{\uparrow}}{2t-E_q}+\frac{1+2\nu_{\downarrow}}{-2t-E_q}=\frac{2}{U},
 \end{equation}
 and that the eigenvectors coefficients $u_q^{\uparrow,\downarrow}$ are related to the pair energies $E_q$ through:
 \begin{equation}
 u_q^{\uparrow}=\frac{1}{2t-E_q} \hspace{.5cm}\text{and}\hspace{.5cm} u_q^{\downarrow}=\frac{1}{-2t-E_q}.
   \end{equation}
\begin{figure}
\centerline{\includegraphics[width=.5\textwidth]{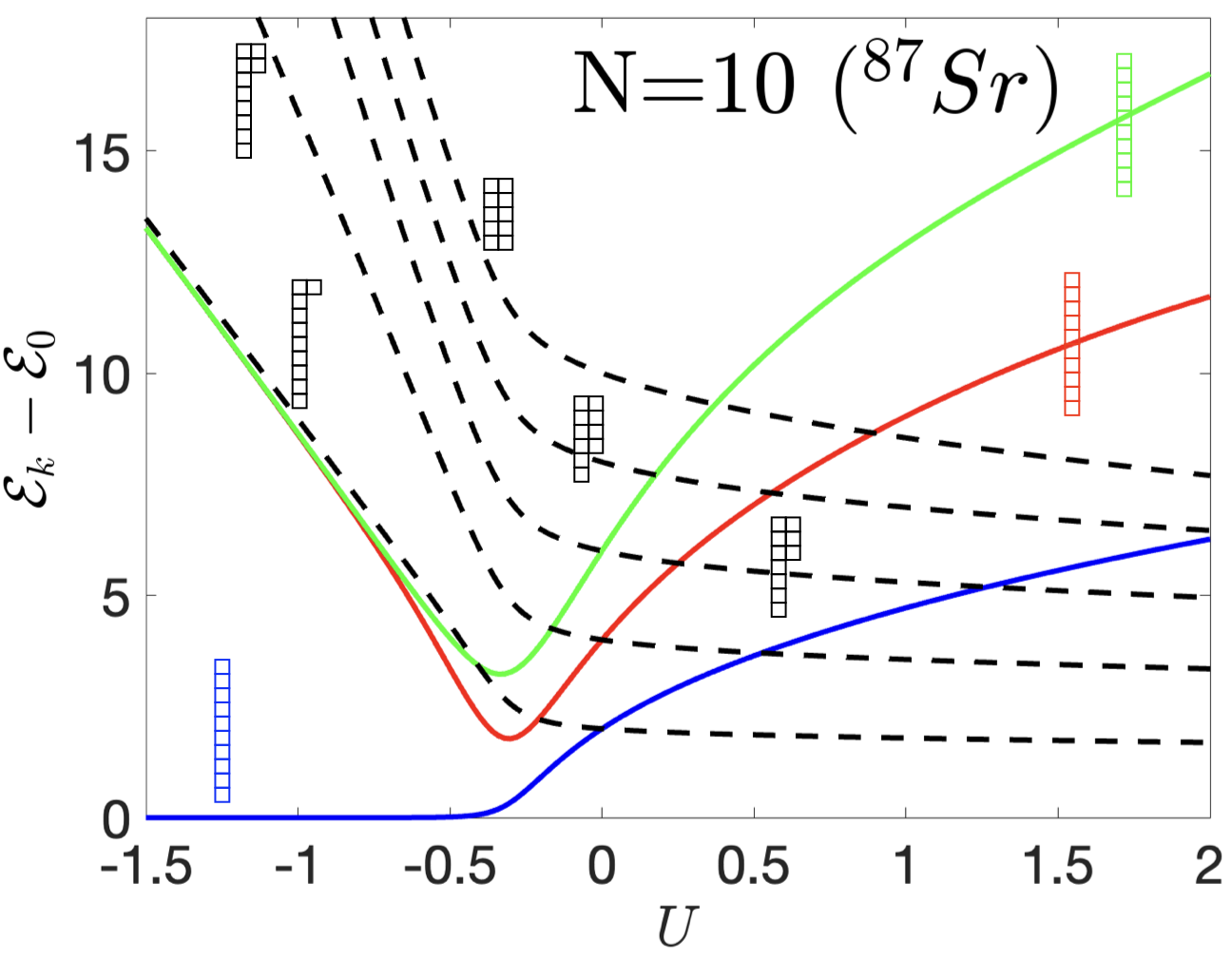}}
\caption{(Color online). Lowest eigen-energies $\mathcal{E}_k$ of the two-sites Fermi-Hubbard Hamiltonian $H$ for $t=1$ as a function of $U$ (cf Eq. \ref{Hamiltonian} in the text) for  $N=10$ and $M=10$ particles. 
The energies  plotted in solid lines (blue for $\mathcal{E}_1$, red for $\mathcal{E}_2$ and green for $\mathcal{E}_3$) are excited energies within the SU(10) singlets irrep $\alpha^{M_1=10}_{M_2=0}\equiv \alpha^{10}_{0}$.
The dashed energies are minimal energies from other irreps.
We have withdrawn the ground state energy $\mathcal{E}_0$ which lives in the SU(10) singlets irrep. The Young Diagram (YD) representations of all the relevant $SU(N)$ irreps are displayed next to the corresponding curves. The first singlet gap $\mathcal{E}_1-\mathcal{E}_0 \equiv \Delta$ tends to zero in the attractive limit exactly as in the second order QPT of the LMG model. }
\label{fig_Energies_SU10}
\end{figure}

In Fig. \ref{fig_Energies_SU10}, we plot the eigen-energies of the Hamiltonian $H$ (cf Eq. \ref{Hamiltonian}) for  $N=10$ (relevant for the cold atoms $^{87} S_r$), given by the solutions of the BAE in Eq. \ref{BAE}, for hopping $t\equiv 1$, as a function of $U$, and for all the different SU(N) irreps at play.
Direct exact diagonalization of the Hamiltonian $H$ (cf Eq. \ref{Hamiltonian}) through the method exposed in \cite{Botzung_2023_PRL} give the very same energies.
We focus here on $M=N=10$ particles, but other situations provide similar spectra.
In particular, for a fixed number of particles $M$, the ground state always lives in the irrep of largest $M_1-M_2$.
The irreps $\alpha^{M_1}_{M_2}$ appear in the Figure \ref{fig_Energies_SU10}, and the corresponding multiplicities are given in Tab. \ref{table: table0}.
As a striking feature, the first excited gap within the $SU(N)$ singlets irrep (which is the fully antisymmetric $N$-boxes one-column irrep, equal here to $\alpha^{10}_{0}$) tends to zero in the attractive $U$ region ($U<0$),
which is reminiscent of the well known second order QPT occurring in the LMG model in the thermodynamical limit \cite{Botet_1982,Botet_1983,Vidal_2004_second}.

\begin{figure}
\includegraphics[width=.5\textwidth]{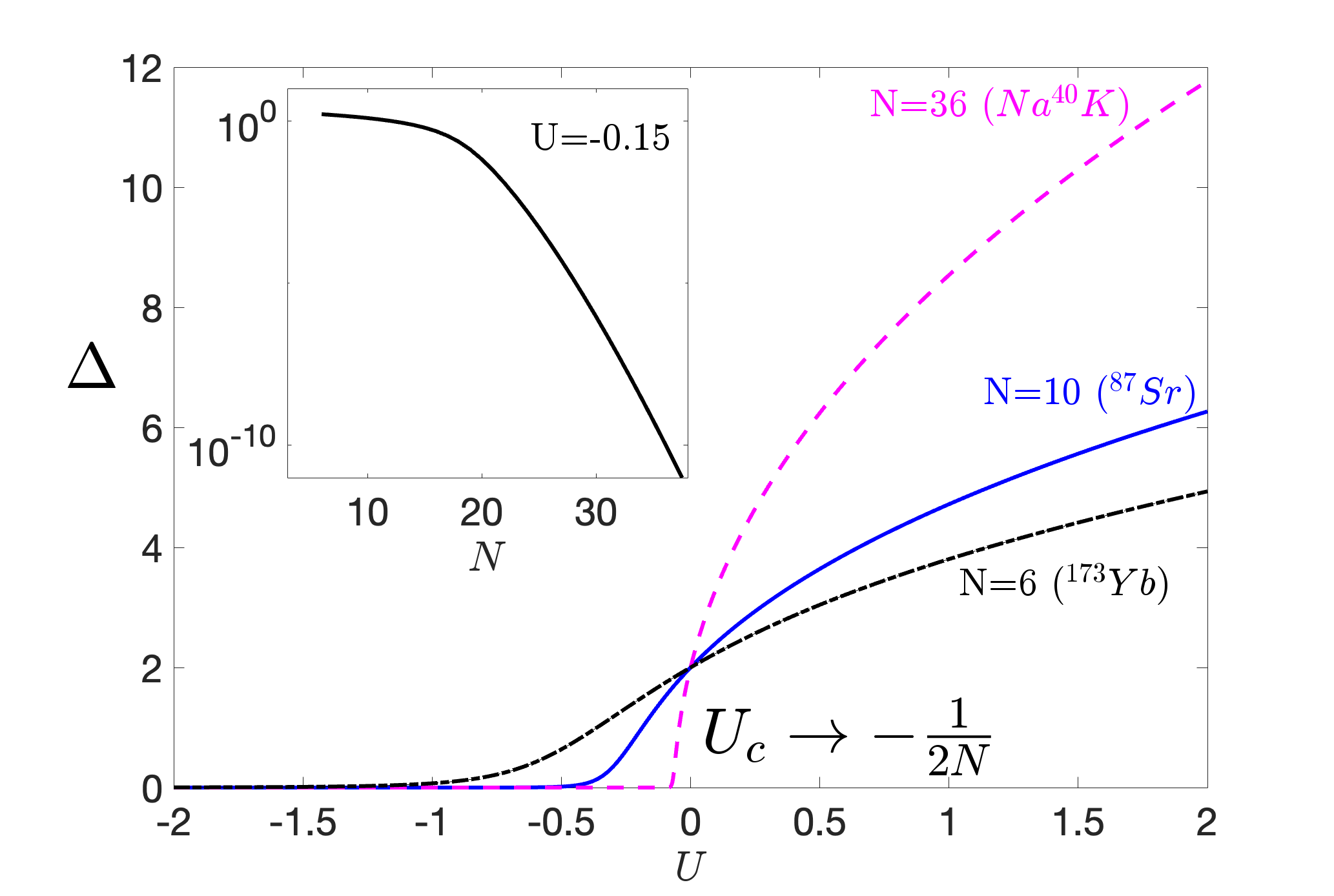}
\caption{Gap $\Delta$ of the Hamiltonian Eq. \ref{Hamiltonian} for $t=1$ as a function of the density-density interaction energy $U$ within the fully anti-symmetric irrep $\alpha^{M_1=N}_{M_2=0}$ (i.e the $SU(N)$ singlets irrep), for various values of $N$. It starts decreasing exponentially with both $\vert U \vert $ and N (as shown in the inset, cf \cite{Newman_1977}), when $U$ is below a (negative) value that converges towards $U_c=-2/N$
for large N. For $U>U_c$, the first gap $\Delta$ behaves as $\omega=\sqrt{2t(2t+UN)}$ (not shown), as expected from the Bogoliubov transformation of the LMG Hamiltonian.}
\label{fig_exponential_gap}
\end{figure}

{\it Second order quantum phase transition.}
For two sites, the Pauli exclusion principle necessitates that the thermodynamic limit (i.e. the number of particles $M \rightarrow+\infty$) must be accompanied by an infinite number of colors ($N \rightarrow+\infty$).
Without loss of generality, we impose $M=N$ particles and we calculate $U_c$, the critical value of $U$ 
at fixed $t=1$ below which the $\mathbb{Z}_2 \equiv exp(i \frac{\pi}{2} (E_{12}+E_{21}))$ parity (with respect to $H$ defined in Eq. \ref{Hamiltonian}) is broken. 
In the thermodynamical limit, the large spins of the LMG model can be replaced by classical spins as in \cite{Botet_1982,Botet_1983}, or equivalently, one can also use the Holstein-Primakoff (HP) \cite{Holstein_Primakoff} transformation like in \cite{Vidal_PRL_2004,Rosensteel_2008}.
Contrary to the Jordan-Schwinger transformation shown in Eq. \ref{Jordan_Schwinger}, the HP transformation for a spin $S$ involves only one bosonic mode and reads: $S_z=S-b^{\dag}b$, $S_{+}=S_{-}^{\dagger}=\sqrt{2S-b^{\dagger}b}\,b$, where $b$ (resp. $b^{\dagger}$) is the annihilation (resp. creation) bosonic operator, and where $2S\equiv M =N$.
Then, the Hamiltonian in Eq. \ref{Hamiltonian_LMG_2} becomes (at leading order in $N$) $H\approx 2 t b^{\dagger}b-\frac{NU}{4}(b-b^{\dagger})^2$,
whose normal eigenmode (obtained through a Bogoliubov transformation) is $\omega=\sqrt{2t(2t+UN)}$, so that for  $t=1$, the critical $U_c$ is:
\begin{equation}
U_c=-\frac{2}{N},
\end{equation}
which goes to $0$ in the asymptotic limit $N= \infty$.
In Fig. \ref{fig_exponential_gap}, we display the first gap $\Delta$
 within the SU($N$) singlets irrep (which is the fully antisymmetric N-boxes one column irrep) for various (experimentally relevant) values of N. By diminishing $U$, the gap drops more abruptly as $N$ increases, revealing important finite-size effects. Actually,
 as shown in the inset, and proved in \cite{Newman_1977}, it decreases exponentially with $N$ for fixed negative $U$.
Moreover, we have checked that the first gap, for positive $U$ behaves as $\omega=\sqrt{2t(2t+UN)}$ (not shown) .

\begin{figure}
\includegraphics[width=.5\textwidth]{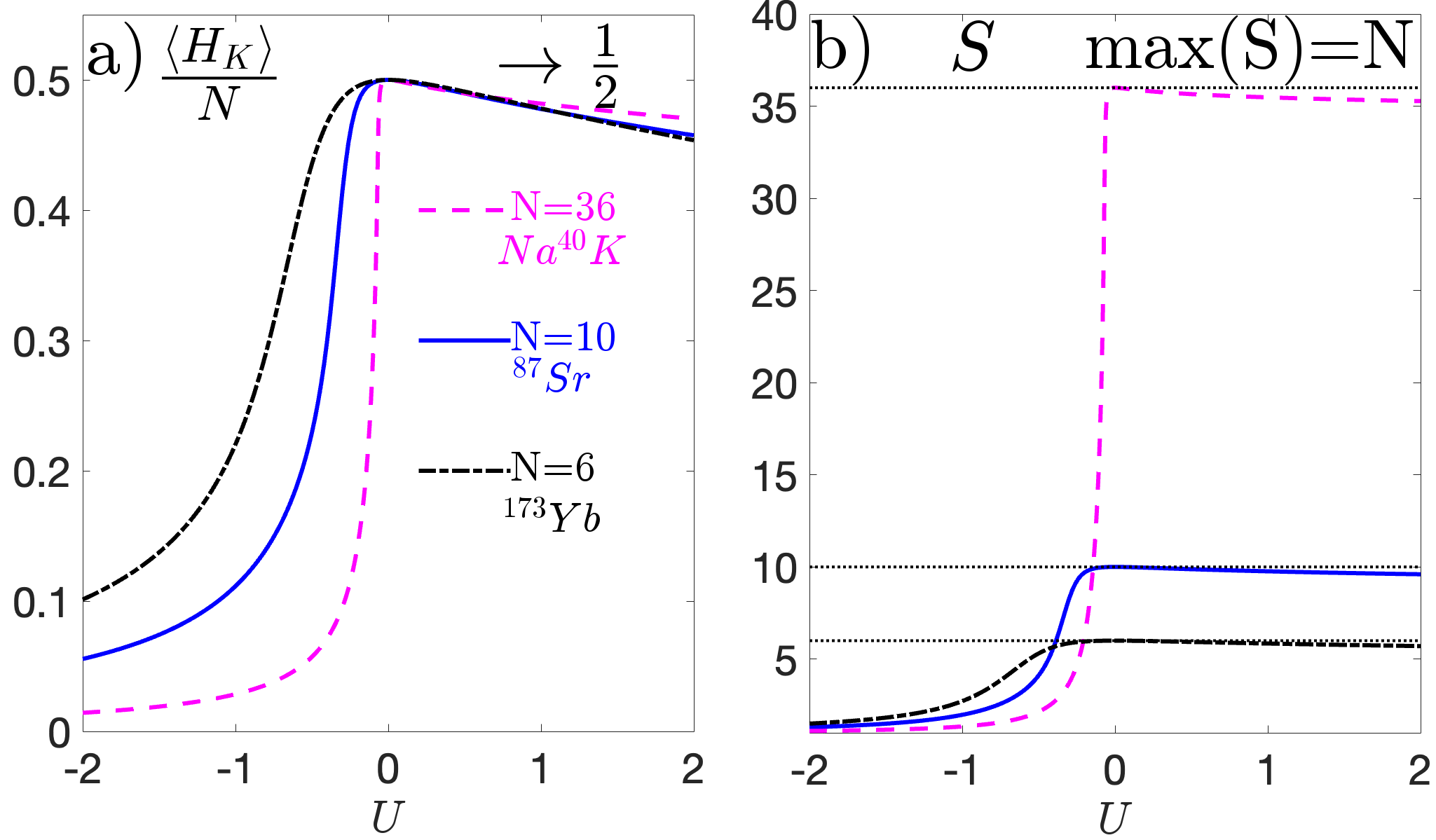}
\caption{a) Expectation value of the kinetic energy $H_K/N\equiv (E_{12}+E_{21})/(2N)$ on the ground state of the two-sites SU(N) Fermi-Hubbard Hamiltonian $H$ defined in Eq. \ref{Hamiltonian} for $M=N$ particles,
for several (experimentally relevant) values of $N$.
From the mapping to the LMG model, it is expected to converge to $1/2$ above the transition, and to behave as $-1/(NU)$ in the symmetry broken phase, as it does.
b) Entanglement Entropy $S$ of the first site with respect to the second. It admits a maximum at $U=0$ where it equals $N$. }
\label{fig_entropy}
\end{figure}

As an order of parameter of the transition, the kinetic energy $H_K\equiv (E_{12}+E_{21})/2$ in Eq. \ref{Hamiltonian} which is equal to the magnetization $S_z$ in Eq. \ref{Hamiltonian_LMG_2}, is expected\cite{Botet_1982,Botet_1983,Vidal_2004_second} to be such that $\langle H_K \rangle/N  =-1/(NU)$ below the transition point and $\langle H_K \rangle/N  =1/2$ above the transition, which is clearly observed in Fig. \ref{fig_entropy} a).

Finally, let's calculate the entanglement entropy which is known as an indicator of the QPT \cite{Osterloh_2002,Osborne_2002,Vidal_2003,Filippone_2011}. 
As opposed to the studies on the LMG model where spins are cut in two (with variable size for each part) \cite{Vidal_2004_first,Vidal_2004_second,Vidal_2007}, it is more physically meaningful here
to consider the entanglement of the first site with respect to the second. Thus, one should consider the first version of the Hamiltonian in Eq. \ref{Hamiltonian}. 
We focus on the ground state for $N=M$ particles, which is a $SU(N)$ singlet, and we use the method developped in \cite{Botzung_2023_PRL,Botzung_2023}  to diagonalize the Hamiltonian on the $(N+1)$ dimensional basis of SSYT of shape $\bar{\alpha}^N_0=[N]$.
We note $\vert k \rangle$ (for $k=0 \cdots N$) the SSYT with $k$ occurrences of the site index $2$ and $N-k$ occurence of the site index $1$.

Then, if efficient for calculating the energies and some observables written as expectation values of the $E_{ij}$ operators ($i,j=1,2$) on the eigenstates, the SSYT basis
should be manipulated carefully when calculating the entanglement since each SSYT already represents a many-body state which is a linear superposition of product states on each site.
For instance, for $N=4$, $\vert 1 \rangle$ will be:
\begin{align}
\label{example_su4_state}
 \ytableausetup{smalltableaux}
\vert 1 \rangle & = \raisebox{-.4ex}{$\ytableaushort{1112}$}=\frac{E_{21}}{2} \raisebox{-.4ex}{$\ytableaushort{1111}$} \nonumber = \frac{E_{21}}{2} \vert 0 \rangle\\
&=\frac{1}{2} \{ c^{\dagger}_{2A} c^{\dagger}_{1B}c^{\dagger}_{1C}c^{\dagger}_{1D}-c^{\dagger}_{2B} c^{\dagger}_{1A}c^{\dagger}_{1C}c^{\dagger}_{1D} \\
&+ c^{\dagger}_{2C} c^{\dagger}_{1A}c^{\dagger}_{1B}c^{\dagger}_{1D}-c^{\dagger}_{2D} c^{\dagger}_{1A}c^{\dagger}_{1B}c^{\dagger}_{1C} \} \vert \vert 0 \rangle \rangle, \nonumber.
\end{align}
where $\vert \vert 0 \rangle \rangle$ is the vacuum, where $\ytableaushort{1111}=\vert 0 \rangle$ is the (unique) SU(4) singlet state on site $1$, and where the $N=4$ flavors are named $A,B,C$ and $D$.
In particular, the factor $1/2$ in the second line of Eq. \ref{example_su4_state} is more generally equal to $1/\sqrt{\binom{N}{k}}$ as there are $\binom{N}{k}$ terms with equal weights in the development of each state $\vert k \rangle$ on the standard basis of product state for each site.
Then, writing the ground state as $\vert G \rangle =\sum_{k=0}^N\beta_k \vert k \rangle $, the partial trace over the states of the second site of the density matrix $\vert G \rangle \langle G \vert$
will be:
\begin{align}
\rho_1=\sum_k \sum_{\bm{\mu}_k} \frac{\vert \beta_k \vert ^2 }{\binom{N}{k}} \vert \bm{\mu}_k \rangle \langle \bm{\mu}_k \vert,
\end{align}
where $\bm{\mu}_k \equiv (\mu_1, \mu_2, \cdots, \mu_k)$ is one of the $\binom{N}{k}$ $k$-uplet of fermionic flavors, so that  
\begin{align}
\label{mu_k}
\vert \bm{\mu}_k \rangle= c^{\dagger}_{1 \mu_1} \cdots c^{\dagger}_{1 \mu_k} \vert \vert 0 \rangle \rangle.
\end{align}

Thus, the entanglement entropy reads:

\begin{align}
S&=-\text{Tr}  \rho_1 \log_2 (\rho_1) = -\sum_{k=0}^N \vert \beta_k \vert ^2 \log_2 \Big{(} \frac{\vert \beta_k \vert ^2}{\binom{N}{k}} \Big{)}.
\end{align}
We have plotted $S$ in Fig. \ref{fig_entropy}. b), 
and it exhibits a maximum for $U=0$, where the analytical calculations become easy: $\beta_k=\sqrt{\binom{N}{k}} \times 2^{-N/2}$ (for $k=0 \cdots N$), so that:
\begin{align}
\label{Max_S}
 {\text{max}}(S)&=S(U=0)=N.
\end{align}
 
This should be put in contrast with the $\log{(N)}$ behavior of the maximum of entropy in the Lieb-Mattis model, the two-level BCS models \cite{Vidal_2007}, or in the two-mode Bose-Einstein condensate \cite{Hines_2003}, where $N$ is a number of particles or spins 1/2.
The origin of this difference is that the $\mathrm{SU}(N)$ FHM2S becomes a "two-modes" model (cf for instance Eqs. \ref{Hamiltonian_LMG} to \ref{Richardson_Hamiltonian}) only under its $\mathrm{SU}(N)$-factorized form, i.e when each $\mathrm{SU}(N)$-symmetric many-body state is mapped onto a spin eigenstate of $S_z$.
But in the standard fermionic $N$-flavors basis (cf last line of Eq. \ref{example_su4_state} or Eq. \ref{mu_k}), the model has $N$ different degenerate orbitals that all contribute to the entanglement entropy,
explaining the behavior shown in Fig. \ref{fig_entropy}. b) and in Eq. \ref{Max_S}. 
 Note that the maximum of the entanglement entropy of some larger $N$-colors systems also scales linearly with $N$\cite{fuhringer2008,Szirmai_2008,Demidio_2015,capponi_phases_2016,nataf_density_2018,shi2024}, with a pre-factor either given by the Calabrese-Cardy formula\cite{calabrese_entanglement_2004} either by an exact Schmidt decomposition \cite{shi2024}, depending on the models.
 
{\it Discussion and Outlook.}
At first,  we were able to give the Richardson BAE for the $\mathrm{SU}(N)$ FHM2S. From a fundamental point of view, it is always useful to give exact results
and to get rid of the apparent exponential complexity (cf Eq. \ref{dimension}) of a many-body problem. 
 
Secondly, guided by the known results on the LMG model, we have shown that this system undergoes a second order QPT for an attractive on-site interaction $U$. For the transition to be observed, one needs to consider systems with a tunable on-site interaction $U$ on a range that contains the critical point (i.e from negative to positive values).
 
 While repulsive on-site interaction has been more frequently reported in  $\mathrm{SU}(N)$ experiments, there are systems where negative interaction can occur.
As a first example, the tunable interactions in a three-state Fermi gas of $^6\text{Li}$ become both $\mathrm{SU}(3)$ symmetric and attractive in the large magnetic field limit as the three two by two pair-wise scattering lengths asymptote to an equal and negative value (See Fig. 1 in \cite{Huckans_2009}).
As for the alkaline-earth atoms $^{173}$Yb and $^{87}$Sr, the absence of electron spin forbids the use of magnetic Feshbach resonance, but inter-orbital Feshbach resonance could be used.
They were already experimentally observed for $^{173}$Yb in \cite{Pagano_2015,Hofer_2015}.
Finally, the proposal based on the ultracold alkali molecules (e.g Na$^{40}$K) is promising, as both the sign and the magnitude of the scattering lengths could be tuned by varying the applied shielding control electric field \cite{Mukherjee_2024,Mukherjee_2024_arxiv}.

In the future, we could search for other phenomena of the LMG model in the  $\mathrm{SU}(N)$ FHM2S. For instance, the LMG model can also host
a first order QPT with a Supersymmetric point \cite{Vidal_2004_first,Unanyan_2003}. It would certainly require some additional fine-tuned interaction between the neighboring sites (i.e of the form $V E_{11}E_{22}$ on top of the other terms in $H$ in Eq. \ref{Hamiltonian}).
Specific experimental protocols for the measurement of the entanglement between site $1$ and site $2$, as well as the finite temperature (T) counterpart of the $T=0$ QPT in the $\mathrm{SU}(N)$ FHM2S could also be investigated.

{\it Acknowledgments}
The author is supported by the IRP EXQMS project from CNRS.

\bibliographystyle{apsrev4-1}
\bibliography{references}

\begin{thebibliography}{68}%
\makeatletter
\providecommand \@ifxundefined [1]{%
 \@ifx{#1\undefined}
}%
\providecommand \@ifnum [1]{%
 \ifnum #1\expandafter \@firstoftwo
 \else \expandafter \@secondoftwo
 \fi
}%
\providecommand \@ifx [1]{%
 \ifx #1\expandafter \@firstoftwo
 \else \expandafter \@secondoftwo
 \fi
}%
\providecommand \natexlab [1]{#1}%
\providecommand \enquote  [1]{``#1''}%
\providecommand \bibnamefont  [1]{#1}%
\providecommand \bibfnamefont [1]{#1}%
\providecommand \citenamefont [1]{#1}%
\providecommand \href@noop [0]{\@secondoftwo}%
\providecommand \href [0]{\begingroup \@sanitize@url \@href}%
\providecommand \@href[1]{\@@startlink{#1}\@@href}%
\providecommand \@@href[1]{\endgroup#1\@@endlink}%
\providecommand \@sanitize@url [0]{\catcode `\\12\catcode `\$12\catcode
  `\&12\catcode `\#12\catcode `\^12\catcode `\_12\catcode `\%12\relax}%
\providecommand \@@startlink[1]{}%
\providecommand \@@endlink[0]{}%
\providecommand \url  [0]{\begingroup\@sanitize@url \@url }%
\providecommand \@url [1]{\endgroup\@href {#1}{\urlprefix }}%
\providecommand \urlprefix  [0]{URL }%
\providecommand \Eprint [0]{\href }%
\providecommand \doibase [0]{http://dx.doi.org/}%
\providecommand \selectlanguage [0]{\@gobble}%
\providecommand \bibinfo  [0]{\@secondoftwo}%
\providecommand \bibfield  [0]{\@secondoftwo}%
\providecommand \translation [1]{[#1]}%
\providecommand \BibitemOpen [0]{}%
\providecommand \bibitemStop [0]{}%
\providecommand \bibitemNoStop [0]{.\EOS\space}%
\providecommand \EOS [0]{\spacefactor3000\relax}%
\providecommand \BibitemShut  [1]{\csname bibitem#1\endcsname}%
\let\auto@bib@innerbib\@empty
\bibitem [{\citenamefont {Hubbard}(1963)}]{Hubbard_1963}%
  \BibitemOpen
  \bibfield  {author} {\bibinfo {author} {\bibfnamefont {J.}~\bibnamefont
  {Hubbard}},\ }\href {\doibase 10.1098/rspa.1963.0204} {\bibfield  {journal}
  {\bibinfo  {journal} {Proceedings of the Royal Society of London. Series A.
  Mathematical and Physical Sciences}\ }\textbf {\bibinfo {volume} {276}},\
  \bibinfo {pages} {238} (\bibinfo {year} {1963})}\BibitemShut {NoStop}%
\bibitem [{\citenamefont {Gutzwiller}(1963)}]{Gutzwiller_1963}%
  \BibitemOpen
  \bibfield  {author} {\bibinfo {author} {\bibfnamefont {M.~C.}\ \bibnamefont
  {Gutzwiller}},\ }\href {\doibase 10.1103/PhysRevLett.10.159} {\bibfield
  {journal} {\bibinfo  {journal} {Phys. Rev. Lett.}\ }\textbf {\bibinfo
  {volume} {10}},\ \bibinfo {pages} {159} (\bibinfo {year} {1963})}\BibitemShut
  {NoStop}%
\bibitem [{\citenamefont {Scalapino}(2012)}]{Scalapino_2012}%
  \BibitemOpen
  \bibfield  {author} {\bibinfo {author} {\bibfnamefont {D.~J.}\ \bibnamefont
  {Scalapino}},\ }\href {\doibase 10.1103/RevModPhys.84.1383} {\bibfield
  {journal} {\bibinfo  {journal} {Rev. Mod. Phys.}\ }\textbf {\bibinfo {volume}
  {84}},\ \bibinfo {pages} {1383} (\bibinfo {year} {2012})}\BibitemShut
  {NoStop}%
\bibitem [{\citenamefont {Anderson}(1987)}]{Anderson_1987}%
  \BibitemOpen
  \bibfield  {author} {\bibinfo {author} {\bibfnamefont {P.~W.}\ \bibnamefont
  {Anderson}},\ }\href {\doibase 10.1126/science.235.4793.1196} {\bibfield
  {journal} {\bibinfo  {journal} {Science}\ }\textbf {\bibinfo {volume}
  {235}},\ \bibinfo {pages} {1196} (\bibinfo {year} {1987})}\BibitemShut
  {NoStop}%
\bibitem [{\citenamefont {Zhang}\ and\ \citenamefont {Rice}(1988)}]{Rice_1988}%
  \BibitemOpen
  \bibfield  {author} {\bibinfo {author} {\bibfnamefont {F.~C.}\ \bibnamefont
  {Zhang}}\ and\ \bibinfo {author} {\bibfnamefont {T.~M.}\ \bibnamefont
  {Rice}},\ }\href {\doibase 10.1103/PhysRevB.37.3759} {\bibfield  {journal}
  {\bibinfo  {journal} {Phys. Rev. B}\ }\textbf {\bibinfo {volume} {37}},\
  \bibinfo {pages} {3759} (\bibinfo {year} {1988})}\BibitemShut {NoStop}%
\bibitem [{\citenamefont {Arovas}\ \emph {et~al.}(2022)\citenamefont {Arovas},
  \citenamefont {Berg}, \citenamefont {Kivelson},\ and\ \citenamefont
  {Raghu}}]{review_Arovas_2022}%
  \BibitemOpen
  \bibfield  {author} {\bibinfo {author} {\bibfnamefont {D.~P.}\ \bibnamefont
  {Arovas}}, \bibinfo {author} {\bibfnamefont {E.}~\bibnamefont {Berg}},
  \bibinfo {author} {\bibfnamefont {S.~A.}\ \bibnamefont {Kivelson}}, \ and\
  \bibinfo {author} {\bibfnamefont {S.}~\bibnamefont {Raghu}},\ }\href
  {\doibase 10.1146/annurev-conmatphys-031620-102024} {\bibfield  {journal}
  {\bibinfo  {journal} {Annual Review of Condensed Matter Physics}\ }\textbf
  {\bibinfo {volume} {13}},\ \bibinfo {pages} {239} (\bibinfo {year}
  {2022})}\BibitemShut {NoStop}%
\bibitem [{\citenamefont {Qin}\ \emph {et~al.}(2022)\citenamefont {Qin},
  \citenamefont {Sch\"{a}fer}, \citenamefont {Andergassen}, \citenamefont
  {Corboz},\ and\ \citenamefont {Gull}}]{review_Corboz_2022}%
  \BibitemOpen
  \bibfield  {author} {\bibinfo {author} {\bibfnamefont {M.}~\bibnamefont
  {Qin}}, \bibinfo {author} {\bibfnamefont {T.}~\bibnamefont {Sch\"{a}fer}},
  \bibinfo {author} {\bibfnamefont {S.}~\bibnamefont {Andergassen}}, \bibinfo
  {author} {\bibfnamefont {P.}~\bibnamefont {Corboz}}, \ and\ \bibinfo {author}
  {\bibfnamefont {E.}~\bibnamefont {Gull}},\ }\href {\doibase
  10.1146/annurev-conmatphys-090921-033948} {\bibfield  {journal} {\bibinfo
  {journal} {Annual Review of Condensed Matter Physics}\ }\textbf {\bibinfo
  {volume} {13}},\ \bibinfo {pages} {275} (\bibinfo {year} {2022})}\BibitemShut
  {NoStop}%
\bibitem [{\citenamefont {Assaraf}\ \emph {et~al.}(1999)\citenamefont
  {Assaraf}, \citenamefont {Azaria}, \citenamefont {Caffarel},\ and\
  \citenamefont {Lecheminant}}]{Assaraf_1999}%
  \BibitemOpen
  \bibfield  {author} {\bibinfo {author} {\bibfnamefont {R.}~\bibnamefont
  {Assaraf}}, \bibinfo {author} {\bibfnamefont {P.}~\bibnamefont {Azaria}},
  \bibinfo {author} {\bibfnamefont {M.}~\bibnamefont {Caffarel}}, \ and\
  \bibinfo {author} {\bibfnamefont {P.}~\bibnamefont {Lecheminant}},\ }\href
  {\doibase 10.1103/PhysRevB.60.2299} {\bibfield  {journal} {\bibinfo
  {journal} {Phys. Rev. B}\ }\textbf {\bibinfo {volume} {60}},\ \bibinfo
  {pages} {2299} (\bibinfo {year} {1999})}\BibitemShut {NoStop}%
\bibitem [{\citenamefont {Honerkamp}\ and\ \citenamefont
  {Hofstetter}(2004)}]{honerkamp_ultrcold_2004}%
  \BibitemOpen
  \bibfield  {author} {\bibinfo {author} {\bibfnamefont {C.}~\bibnamefont
  {Honerkamp}}\ and\ \bibinfo {author} {\bibfnamefont {W.}~\bibnamefont
  {Hofstetter}},\ }\href
  {https://link.aps.org/doi/10.1103/PhysRevLett.92.170403} {\bibfield
  {journal} {\bibinfo  {journal} {Phys. Rev. Lett.}\ }\textbf {\bibinfo
  {volume} {92}},\ \bibinfo {pages} {170403} (\bibinfo {year}
  {2004})}\BibitemShut {NoStop}%
\bibitem [{\citenamefont {Capponi}\ \emph {et~al.}(2016)\citenamefont
  {Capponi}, \citenamefont {Lecheminant},\ and\ \citenamefont
  {Totsuka}}]{capponi_phases_2016}%
  \BibitemOpen
  \bibfield  {author} {\bibinfo {author} {\bibfnamefont {S.}~\bibnamefont
  {Capponi}}, \bibinfo {author} {\bibfnamefont {P.}~\bibnamefont
  {Lecheminant}}, \ and\ \bibinfo {author} {\bibfnamefont {K.}~\bibnamefont
  {Totsuka}},\ }\href
  {http://www.sciencedirect.com/science/article/pii/S0003491616000130}
  {\bibfield  {journal} {\bibinfo  {journal} {Annals of Physics}\ }\textbf
  {\bibinfo {volume} {367}},\ \bibinfo {pages} {50 } (\bibinfo {year}
  {2016})}\BibitemShut {NoStop}%
\bibitem [{\citenamefont {Ibarra-Garcia-Padilla}\ and\ \citenamefont
  {Choudhury}(2024)}]{Ibarra_Garcia_Padilla_2024}%
  \BibitemOpen
  \bibfield  {author} {\bibinfo {author} {\bibfnamefont {E.}~\bibnamefont
  {Ibarra-Garcia-Padilla}}\ and\ \bibinfo {author} {\bibfnamefont
  {S.}~\bibnamefont {Choudhury}},\ }\href {\doibase 10.1088/1361-648X/ad9658}
  {\bibfield  {journal} {\bibinfo  {journal} {Journal of Physics: Condensed
  Matter}\ }\textbf {\bibinfo {volume} {37}},\ \bibinfo {pages} {083003}
  (\bibinfo {year} {2024})}\BibitemShut {NoStop}%
\bibitem [{\citenamefont {Affleck}(1986)}]{affleck_exact_1986}%
  \BibitemOpen
  \bibfield  {author} {\bibinfo {author} {\bibfnamefont {I.}~\bibnamefont
  {Affleck}},\ }\href
  {http://www.sciencedirect.com/science/article/pii/0550321386901677}
  {\bibfield  {journal} {\bibinfo  {journal} {Nuclear Physics B}\ }\textbf
  {\bibinfo {volume} {265}},\ \bibinfo {pages} {409 } (\bibinfo {year}
  {1986})}\BibitemShut {NoStop}%
\bibitem [{\citenamefont {Affleck}\ and\ \citenamefont
  {Marston}(1988)}]{Affleck_1988}%
  \BibitemOpen
  \bibfield  {author} {\bibinfo {author} {\bibfnamefont {I.}~\bibnamefont
  {Affleck}}\ and\ \bibinfo {author} {\bibfnamefont {J.~B.}\ \bibnamefont
  {Marston}},\ }\href {\doibase 10.1103/PhysRevB.37.3774} {\bibfield  {journal}
  {\bibinfo  {journal} {Phys. Rev. B}\ }\textbf {\bibinfo {volume} {37}},\
  \bibinfo {pages} {3774} (\bibinfo {year} {1988})}\BibitemShut {NoStop}%
\bibitem [{\citenamefont {Rokhsar}(1990)}]{Rokhsar_1990}%
  \BibitemOpen
  \bibfield  {author} {\bibinfo {author} {\bibfnamefont {D.~S.}\ \bibnamefont
  {Rokhsar}},\ }\href {\doibase 10.1103/PhysRevB.42.2526} {\bibfield  {journal}
  {\bibinfo  {journal} {Phys. Rev. B}\ }\textbf {\bibinfo {volume} {42}},\
  \bibinfo {pages} {2526} (\bibinfo {year} {1990})}\BibitemShut {NoStop}%
\bibitem [{\citenamefont {Marder}\ \emph {et~al.}(1990)\citenamefont {Marder},
  \citenamefont {Papanicolaou},\ and\ \citenamefont {Psaltakis}}]{Marder_1990}%
  \BibitemOpen
  \bibfield  {author} {\bibinfo {author} {\bibfnamefont {M.}~\bibnamefont
  {Marder}}, \bibinfo {author} {\bibfnamefont {N.}~\bibnamefont
  {Papanicolaou}}, \ and\ \bibinfo {author} {\bibfnamefont {G.~C.}\
  \bibnamefont {Psaltakis}},\ }\href {\doibase 10.1103/PhysRevB.41.6920}
  {\bibfield  {journal} {\bibinfo  {journal} {Phys. Rev. B}\ }\textbf {\bibinfo
  {volume} {41}},\ \bibinfo {pages} {6920} (\bibinfo {year}
  {1990})}\BibitemShut {NoStop}%
\bibitem [{\citenamefont {Pati}\ \emph {et~al.}(1998)\citenamefont {Pati},
  \citenamefont {Singh},\ and\ \citenamefont {Khomskii}}]{Khomskii_1998}%
  \BibitemOpen
  \bibfield  {author} {\bibinfo {author} {\bibfnamefont {S.~K.}\ \bibnamefont
  {Pati}}, \bibinfo {author} {\bibfnamefont {R.~R.~P.}\ \bibnamefont {Singh}},
  \ and\ \bibinfo {author} {\bibfnamefont {D.~I.}\ \bibnamefont {Khomskii}},\
  }\href {\doibase 10.1103/PhysRevLett.81.5406} {\bibfield  {journal} {\bibinfo
   {journal} {Phys. Rev. Lett.}\ }\textbf {\bibinfo {volume} {81}},\ \bibinfo
  {pages} {5406} (\bibinfo {year} {1998})}\BibitemShut {NoStop}%
\bibitem [{\citenamefont {Yamada}\ \emph {et~al.}(2018)\citenamefont {Yamada},
  \citenamefont {Oshikawa},\ and\ \citenamefont {Jackeli}}]{Yamada_2018}%
  \BibitemOpen
  \bibfield  {author} {\bibinfo {author} {\bibfnamefont {M.~G.}\ \bibnamefont
  {Yamada}}, \bibinfo {author} {\bibfnamefont {M.}~\bibnamefont {Oshikawa}}, \
  and\ \bibinfo {author} {\bibfnamefont {G.}~\bibnamefont {Jackeli}},\ }\href
  {\doibase 10.1103/PhysRevLett.121.097201} {\bibfield  {journal} {\bibinfo
  {journal} {Phys. Rev. Lett.}\ }\textbf {\bibinfo {volume} {121}},\ \bibinfo
  {pages} {097201} (\bibinfo {year} {2018})}\BibitemShut {NoStop}%
\bibitem [{\citenamefont {Zhang}\ \emph {et~al.}(2021)\citenamefont {Zhang},
  \citenamefont {Sheng},\ and\ \citenamefont {Vishwanath}}]{zhang2021}%
  \BibitemOpen
  \bibfield  {author} {\bibinfo {author} {\bibfnamefont {Y.-H.}\ \bibnamefont
  {Zhang}}, \bibinfo {author} {\bibfnamefont {D.~N.}\ \bibnamefont {Sheng}}, \
  and\ \bibinfo {author} {\bibfnamefont {A.}~\bibnamefont {Vishwanath}},\
  }\href {\doibase 10.1103/PhysRevLett.127.247701} {\bibfield  {journal}
  {\bibinfo  {journal} {Phys. Rev. Lett.}\ }\textbf {\bibinfo {volume} {127}},\
  \bibinfo {pages} {247701} (\bibinfo {year} {2021})}\BibitemShut {NoStop}%
\bibitem [{\citenamefont {Wu}\ \emph {et~al.}(2003)\citenamefont {Wu},
  \citenamefont {Hu},\ and\ \citenamefont {Zhang}}]{wu_exact_2003}%
  \BibitemOpen
  \bibfield  {author} {\bibinfo {author} {\bibfnamefont {C.}~\bibnamefont
  {Wu}}, \bibinfo {author} {\bibfnamefont {J.-p.}\ \bibnamefont {Hu}}, \ and\
  \bibinfo {author} {\bibfnamefont {S.-c.}\ \bibnamefont {Zhang}},\ }\href
  {https://link.aps.org/doi/10.1103/PhysRevLett.91.186402} {\bibfield
  {journal} {\bibinfo  {journal} {Phys. Rev. Lett.}\ }\textbf {\bibinfo
  {volume} {91}},\ \bibinfo {pages} {186402} (\bibinfo {year}
  {2003})}\BibitemShut {NoStop}%
\bibitem [{\citenamefont {Wu}(2006)}]{Wu_review_2006}%
  \BibitemOpen
  \bibfield  {author} {\bibinfo {author} {\bibfnamefont {C.}~\bibnamefont
  {Wu}},\ }\href {\doibase 10.1142/S0217984906012213} {\bibfield  {journal}
  {\bibinfo  {journal} {Modern Physics Letters B}\ }\textbf {\bibinfo {volume}
  {20}},\ \bibinfo {pages} {1707} (\bibinfo {year} {2006})}\BibitemShut
  {NoStop}%
\bibitem [{\citenamefont {Gorshkov}\ \emph {et~al.}(2010)\citenamefont
  {Gorshkov}, \citenamefont {Hermele}, \citenamefont {Gurarie}, \citenamefont
  {Xu}, \citenamefont {Julienne}, \citenamefont {Ye}, \citenamefont {Zoller},
  \citenamefont {Demler}, \citenamefont {Lukin},\ and\ \citenamefont
  {Rey}}]{gorshkov_two_2010}%
  \BibitemOpen
  \bibfield  {author} {\bibinfo {author} {\bibfnamefont {A.~V.}\ \bibnamefont
  {Gorshkov}}, \bibinfo {author} {\bibfnamefont {M.}~\bibnamefont {Hermele}},
  \bibinfo {author} {\bibfnamefont {V.}~\bibnamefont {Gurarie}}, \bibinfo
  {author} {\bibfnamefont {C.}~\bibnamefont {Xu}}, \bibinfo {author}
  {\bibfnamefont {P.~S.}\ \bibnamefont {Julienne}}, \bibinfo {author}
  {\bibfnamefont {J.}~\bibnamefont {Ye}}, \bibinfo {author} {\bibfnamefont
  {P.}~\bibnamefont {Zoller}}, \bibinfo {author} {\bibfnamefont
  {E.}~\bibnamefont {Demler}}, \bibinfo {author} {\bibfnamefont {M.~D.}\
  \bibnamefont {Lukin}}, \ and\ \bibinfo {author} {\bibfnamefont
  {A.}~\bibnamefont {Rey}},\ }\href {https://www.nature.com/articles/nphys1535}
  {\bibfield  {journal} {\bibinfo  {journal} {Nature physics}\ }\textbf
  {\bibinfo {volume} {6}},\ \bibinfo {pages} {289} (\bibinfo {year}
  {2010})}\BibitemShut {NoStop}%
\bibitem [{\citenamefont {Cazalilla}\ and\ \citenamefont
  {Rey}(2014)}]{Cazalilla_2014}%
  \BibitemOpen
  \bibfield  {author} {\bibinfo {author} {\bibfnamefont {M.~A.}\ \bibnamefont
  {Cazalilla}}\ and\ \bibinfo {author} {\bibfnamefont {A.~M.}\ \bibnamefont
  {Rey}},\ }\href {\doibase 10.1088/0034-4885/77/12/124401} {\bibfield
  {journal} {\bibinfo  {journal} {Reports on Progress in Physics}\ }\textbf
  {\bibinfo {volume} {77}},\ \bibinfo {pages} {124401} (\bibinfo {year}
  {2014})}\BibitemShut {NoStop}%
\bibitem [{\citenamefont {Taie}\ \emph {et~al.}(2010)\citenamefont {Taie},
  \citenamefont {Takasu}, \citenamefont {Sugawa}, \citenamefont {Yamazaki},
  \citenamefont {Tsujimoto}, \citenamefont {Murakami},\ and\ \citenamefont
  {Takahashi}}]{taie_realization_2010}%
  \BibitemOpen
  \bibfield  {author} {\bibinfo {author} {\bibfnamefont {S.}~\bibnamefont
  {Taie}}, \bibinfo {author} {\bibfnamefont {Y.}~\bibnamefont {Takasu}},
  \bibinfo {author} {\bibfnamefont {S.}~\bibnamefont {Sugawa}}, \bibinfo
  {author} {\bibfnamefont {R.}~\bibnamefont {Yamazaki}}, \bibinfo {author}
  {\bibfnamefont {T.}~\bibnamefont {Tsujimoto}}, \bibinfo {author}
  {\bibfnamefont {R.}~\bibnamefont {Murakami}}, \ and\ \bibinfo {author}
  {\bibfnamefont {Y.}~\bibnamefont {Takahashi}},\ }\href
  {https://link.aps.org/doi/10.1103/PhysRevLett.105.190401} {\bibfield
  {journal} {\bibinfo  {journal} {Phys. Rev. Lett.}\ }\textbf {\bibinfo
  {volume} {105}},\ \bibinfo {pages} {190401} (\bibinfo {year}
  {2010})}\BibitemShut {NoStop}%
\bibitem [{\citenamefont {DeSalvo}\ \emph {et~al.}(2010)\citenamefont
  {DeSalvo}, \citenamefont {Yan}, \citenamefont {Mickelson}, \citenamefont
  {Martinez~de Escobar},\ and\ \citenamefont
  {Killian}}]{desalvo_degenerate_2010}%
  \BibitemOpen
  \bibfield  {author} {\bibinfo {author} {\bibfnamefont {B.~J.}\ \bibnamefont
  {DeSalvo}}, \bibinfo {author} {\bibfnamefont {M.}~\bibnamefont {Yan}},
  \bibinfo {author} {\bibfnamefont {P.~G.}\ \bibnamefont {Mickelson}}, \bibinfo
  {author} {\bibfnamefont {Y.~N.}\ \bibnamefont {Martinez~de Escobar}}, \ and\
  \bibinfo {author} {\bibfnamefont {T.~C.}\ \bibnamefont {Killian}},\ }\href
  {https://link.aps.org/doi/10.1103/PhysRevLett.105.030402} {\bibfield
  {journal} {\bibinfo  {journal} {Phys. Rev. Lett.}\ }\textbf {\bibinfo
  {volume} {105}},\ \bibinfo {pages} {030402} (\bibinfo {year}
  {2010})}\BibitemShut {NoStop}%
\bibitem [{\citenamefont {Zhang}\ \emph {et~al.}(2014)\citenamefont {Zhang},
  \citenamefont {Bishof}, \citenamefont {Bromley}, \citenamefont {Kraus},
  \citenamefont {Safronova}, \citenamefont {Zoller}, \citenamefont {Rey},\ and\
  \citenamefont {Ye}}]{zhang_spectroscopic_2014}%
  \BibitemOpen
  \bibfield  {author} {\bibinfo {author} {\bibfnamefont {X.}~\bibnamefont
  {Zhang}}, \bibinfo {author} {\bibfnamefont {M.}~\bibnamefont {Bishof}},
  \bibinfo {author} {\bibfnamefont {S.~L.}\ \bibnamefont {Bromley}}, \bibinfo
  {author} {\bibfnamefont {C.~V.}\ \bibnamefont {Kraus}}, \bibinfo {author}
  {\bibfnamefont {M.~S.}\ \bibnamefont {Safronova}}, \bibinfo {author}
  {\bibfnamefont {P.}~\bibnamefont {Zoller}}, \bibinfo {author} {\bibfnamefont
  {A.~M.}\ \bibnamefont {Rey}}, \ and\ \bibinfo {author} {\bibfnamefont
  {J.}~\bibnamefont {Ye}},\ }\href {\doibase 10.1126/science.1254978}
  {\bibfield  {journal} {\bibinfo  {journal} {Science}\ }\textbf {\bibinfo
  {volume} {345}},\ \bibinfo {pages} {1467} (\bibinfo {year}
  {2014})}\BibitemShut {NoStop}%
\bibitem [{\citenamefont {Pagano}\ \emph {et~al.}(2014)\citenamefont {Pagano},
  \citenamefont {Mancini}, \citenamefont {Cappellini}, \citenamefont
  {Lombardi}, \citenamefont {Sch{\"a}fer}, \citenamefont {Hu}, \citenamefont
  {Liu}, \citenamefont {Catani}, \citenamefont {Sias}, \citenamefont {Inguscio}
  \emph {et~al.}}]{pagano_one_2014}%
  \BibitemOpen
  \bibfield  {author} {\bibinfo {author} {\bibfnamefont {G.}~\bibnamefont
  {Pagano}}, \bibinfo {author} {\bibfnamefont {M.}~\bibnamefont {Mancini}},
  \bibinfo {author} {\bibfnamefont {G.}~\bibnamefont {Cappellini}}, \bibinfo
  {author} {\bibfnamefont {P.}~\bibnamefont {Lombardi}}, \bibinfo {author}
  {\bibfnamefont {F.}~\bibnamefont {Sch{\"a}fer}}, \bibinfo {author}
  {\bibfnamefont {H.}~\bibnamefont {Hu}}, \bibinfo {author} {\bibfnamefont
  {X.-J.}\ \bibnamefont {Liu}}, \bibinfo {author} {\bibfnamefont
  {J.}~\bibnamefont {Catani}}, \bibinfo {author} {\bibfnamefont
  {C.}~\bibnamefont {Sias}}, \bibinfo {author} {\bibfnamefont {M.}~\bibnamefont
  {Inguscio}},  \emph {et~al.},\ }\href
  {https://www.nature.com/articles/nphys2878} {\bibfield  {journal} {\bibinfo
  {journal} {Nature Physics}\ }\textbf {\bibinfo {volume} {10}},\ \bibinfo
  {pages} {198} (\bibinfo {year} {2014})}\BibitemShut {NoStop}%
\bibitem [{\citenamefont {Taie}\ \emph {et~al.}(2012)\citenamefont {Taie},
  \citenamefont {Yamazaki}, \citenamefont {Sugawa},\ and\ \citenamefont
  {Takahashi}}]{taie_su6_2012}%
  \BibitemOpen
  \bibfield  {author} {\bibinfo {author} {\bibfnamefont {S.}~\bibnamefont
  {Taie}}, \bibinfo {author} {\bibfnamefont {R.}~\bibnamefont {Yamazaki}},
  \bibinfo {author} {\bibfnamefont {S.}~\bibnamefont {Sugawa}}, \ and\ \bibinfo
  {author} {\bibfnamefont {Y.}~\bibnamefont {Takahashi}},\ }\href
  {https://www.nature.com/articles/nphys2430} {\bibfield  {journal} {\bibinfo
  {journal} {Nature Physics}\ }\textbf {\bibinfo {volume} {8}},\ \bibinfo
  {pages} {825} (\bibinfo {year} {2012})}\BibitemShut {NoStop}%
\bibitem [{\citenamefont {Abeln}\ \emph {et~al.}(2021)\citenamefont {Abeln},
  \citenamefont {Sponselee}, \citenamefont {Diem}, \citenamefont {Pintul},
  \citenamefont {Sengstock},\ and\ \citenamefont {Becker}}]{Becker_2021}%
  \BibitemOpen
  \bibfield  {author} {\bibinfo {author} {\bibfnamefont {B.}~\bibnamefont
  {Abeln}}, \bibinfo {author} {\bibfnamefont {K.}~\bibnamefont {Sponselee}},
  \bibinfo {author} {\bibfnamefont {M.}~\bibnamefont {Diem}}, \bibinfo {author}
  {\bibfnamefont {N.}~\bibnamefont {Pintul}}, \bibinfo {author} {\bibfnamefont
  {K.}~\bibnamefont {Sengstock}}, \ and\ \bibinfo {author} {\bibfnamefont
  {C.}~\bibnamefont {Becker}},\ }\href {\doibase 10.1103/PhysRevA.103.033315}
  {\bibfield  {journal} {\bibinfo  {journal} {Phys. Rev. A}\ }\textbf {\bibinfo
  {volume} {103}},\ \bibinfo {pages} {033315} (\bibinfo {year}
  {2021})}\BibitemShut {NoStop}%
\bibitem [{\citenamefont {Tusi}\ \emph {et~al.}(2022)\citenamefont {Tusi},
  \citenamefont {Franchi}, \citenamefont {Livi}, \citenamefont {Baumann},
  \citenamefont {Benedicto~Orenes}, \citenamefont {Del~Re}, \citenamefont
  {Barfknecht}, \citenamefont {Zhou}, \citenamefont {Inguscio}, \citenamefont
  {Cappellini}, \citenamefont {Capone}, \citenamefont {Catani},\ and\
  \citenamefont {Fallani}}]{Fallani_2022}%
  \BibitemOpen
  \bibfield  {author} {\bibinfo {author} {\bibfnamefont {D.}~\bibnamefont
  {Tusi}}, \bibinfo {author} {\bibfnamefont {L.}~\bibnamefont {Franchi}},
  \bibinfo {author} {\bibfnamefont {L.~F.}\ \bibnamefont {Livi}}, \bibinfo
  {author} {\bibfnamefont {K.}~\bibnamefont {Baumann}}, \bibinfo {author}
  {\bibfnamefont {D.}~\bibnamefont {Benedicto~Orenes}}, \bibinfo {author}
  {\bibfnamefont {L.}~\bibnamefont {Del~Re}}, \bibinfo {author} {\bibfnamefont
  {R.~E.}\ \bibnamefont {Barfknecht}}, \bibinfo {author} {\bibfnamefont
  {T.~W.}\ \bibnamefont {Zhou}}, \bibinfo {author} {\bibfnamefont
  {M.}~\bibnamefont {Inguscio}}, \bibinfo {author} {\bibfnamefont
  {G.}~\bibnamefont {Cappellini}}, \bibinfo {author} {\bibfnamefont
  {M.}~\bibnamefont {Capone}}, \bibinfo {author} {\bibfnamefont
  {J.}~\bibnamefont {Catani}}, \ and\ \bibinfo {author} {\bibfnamefont
  {L.}~\bibnamefont {Fallani}},\ }\href {\doibase 10.1038/s41567-022-01726-5}
  {\bibfield  {journal} {\bibinfo  {journal} {Nature Physics}\ }\textbf
  {\bibinfo {volume} {18}},\ \bibinfo {pages} {1201} (\bibinfo {year}
  {2022})}\BibitemShut {NoStop}%
\bibitem [{\citenamefont {Hofrichter}\ \emph {et~al.}(2016)\citenamefont
  {Hofrichter}, \citenamefont {Riegger}, \citenamefont {Scazza}, \citenamefont
  {H\"ofer}, \citenamefont {Fernandes}, \citenamefont {Bloch},\ and\
  \citenamefont {F\"olling}}]{hofrichter_direct_2016}%
  \BibitemOpen
  \bibfield  {author} {\bibinfo {author} {\bibfnamefont {C.}~\bibnamefont
  {Hofrichter}}, \bibinfo {author} {\bibfnamefont {L.}~\bibnamefont {Riegger}},
  \bibinfo {author} {\bibfnamefont {F.}~\bibnamefont {Scazza}}, \bibinfo
  {author} {\bibfnamefont {M.}~\bibnamefont {H\"ofer}}, \bibinfo {author}
  {\bibfnamefont {D.~R.}\ \bibnamefont {Fernandes}}, \bibinfo {author}
  {\bibfnamefont {I.}~\bibnamefont {Bloch}}, \ and\ \bibinfo {author}
  {\bibfnamefont {S.}~\bibnamefont {F\"olling}},\ }\href {\doibase
  10.1103/PhysRevX.6.021030} {\bibfield  {journal} {\bibinfo  {journal} {Phys.
  Rev. X}\ }\textbf {\bibinfo {volume} {6}},\ \bibinfo {pages} {021030}
  (\bibinfo {year} {2016})}\BibitemShut {NoStop}%
\bibitem [{\citenamefont {Taie}\ \emph {et~al.}(2022)\citenamefont {Taie},
  \citenamefont {Ibarra-Garc{\'\i}a-Padilla}, \citenamefont {Nishizawa},
  \citenamefont {Takasu}, \citenamefont {Kuno}, \citenamefont {Wei},
  \citenamefont {Scalettar}, \citenamefont {Hazzard},\ and\ \citenamefont
  {Takahashi}}]{taie2020observation}%
  \BibitemOpen
  \bibfield  {author} {\bibinfo {author} {\bibfnamefont {S.}~\bibnamefont
  {Taie}}, \bibinfo {author} {\bibfnamefont {E.}~\bibnamefont
  {Ibarra-Garc{\'\i}a-Padilla}}, \bibinfo {author} {\bibfnamefont
  {N.}~\bibnamefont {Nishizawa}}, \bibinfo {author} {\bibfnamefont
  {Y.}~\bibnamefont {Takasu}}, \bibinfo {author} {\bibfnamefont
  {Y.}~\bibnamefont {Kuno}}, \bibinfo {author} {\bibfnamefont {H.-T.}\
  \bibnamefont {Wei}}, \bibinfo {author} {\bibfnamefont {R.~T.}\ \bibnamefont
  {Scalettar}}, \bibinfo {author} {\bibfnamefont {K.~R.~A.}\ \bibnamefont
  {Hazzard}}, \ and\ \bibinfo {author} {\bibfnamefont {Y.}~\bibnamefont
  {Takahashi}},\ }\href {\doibase 10.1038/s41567-022-01725-6} {\bibfield
  {journal} {\bibinfo  {journal} {Nature Physics}\ }\textbf {\bibinfo {volume}
  {18}},\ \bibinfo {pages} {1356} (\bibinfo {year} {2022})}\BibitemShut
  {NoStop}%
\bibitem [{\citenamefont {Pasqualetti}\ \emph {et~al.}(2024)\citenamefont
  {Pasqualetti}, \citenamefont {Bettermann}, \citenamefont {Darkwah~Oppong},
  \citenamefont {Ibarra-Garc\'{\i}a-Padilla}, \citenamefont {Dasgupta},
  \citenamefont {Scalettar}, \citenamefont {Hazzard}, \citenamefont {Bloch},\
  and\ \citenamefont {F\"olling}}]{Pasqualetti_2024}%
  \BibitemOpen
  \bibfield  {author} {\bibinfo {author} {\bibfnamefont {G.}~\bibnamefont
  {Pasqualetti}}, \bibinfo {author} {\bibfnamefont {O.}~\bibnamefont
  {Bettermann}}, \bibinfo {author} {\bibfnamefont {N.}~\bibnamefont
  {Darkwah~Oppong}}, \bibinfo {author} {\bibfnamefont {E.}~\bibnamefont
  {Ibarra-Garc\'{\i}a-Padilla}}, \bibinfo {author} {\bibfnamefont
  {S.}~\bibnamefont {Dasgupta}}, \bibinfo {author} {\bibfnamefont {R.~T.}\
  \bibnamefont {Scalettar}}, \bibinfo {author} {\bibfnamefont {K.~R.~A.}\
  \bibnamefont {Hazzard}}, \bibinfo {author} {\bibfnamefont {I.}~\bibnamefont
  {Bloch}}, \ and\ \bibinfo {author} {\bibfnamefont {S.}~\bibnamefont
  {F\"olling}},\ }\href {\doibase 10.1103/PhysRevLett.132.083401} {\bibfield
  {journal} {\bibinfo  {journal} {Phys. Rev. Lett.}\ }\textbf {\bibinfo
  {volume} {132}},\ \bibinfo {pages} {083401} (\bibinfo {year}
  {2024})}\BibitemShut {NoStop}%
\bibitem [{\citenamefont {Richardson}(1968)}]{Richardson}%
  \BibitemOpen
  \bibfield  {author} {\bibinfo {author} {\bibfnamefont {R.~W.}\ \bibnamefont
  {Richardson}},\ }\href {\doibase 10.1063/1.1664719} {\bibfield  {journal}
  {\bibinfo  {journal} {Journal of Mathematical Physics}\ }\textbf {\bibinfo
  {volume} {9}},\ \bibinfo {pages} {1327} (\bibinfo {year} {1968})},\ \Eprint
  {http://arxiv.org/abs/https://pubs.aip.org/aip/jmp/article-pdf/9/9/1327/19089387/1327\_1\_online.pdf}
  {https://pubs.aip.org/aip/jmp/article-pdf/9/9/1327/19089387/1327\_1\_online.pdf}
  \BibitemShut {NoStop}%
\bibitem [{\citenamefont {Lipkin}\ \emph {et~al.}(1965)\citenamefont {Lipkin},
  \citenamefont {Meshkov},\ and\ \citenamefont {Glick}}]{LMG_1965}%
  \BibitemOpen
  \bibfield  {author} {\bibinfo {author} {\bibfnamefont {H.}~\bibnamefont
  {Lipkin}}, \bibinfo {author} {\bibfnamefont {N.}~\bibnamefont {Meshkov}}, \
  and\ \bibinfo {author} {\bibfnamefont {A.}~\bibnamefont {Glick}},\ }\href
  {\doibase https://doi.org/10.1016/0029-5582(65)90862-X} {\bibfield  {journal}
  {\bibinfo  {journal} {Nuclear Physics}\ }\textbf {\bibinfo {volume} {62}},\
  \bibinfo {pages} {188} (\bibinfo {year} {1965})}\BibitemShut {NoStop}%
\bibitem [{\citenamefont {Mukherjee}\ \emph {et~al.}(2024)\citenamefont
  {Mukherjee}, \citenamefont {Hutson},\ and\ \citenamefont
  {Hazzard}}]{Mukherjee_2024}%
  \BibitemOpen
  \bibfield  {author} {\bibinfo {author} {\bibfnamefont {B.}~\bibnamefont
  {Mukherjee}}, \bibinfo {author} {\bibfnamefont {J.~M.}\ \bibnamefont
  {Hutson}}, \ and\ \bibinfo {author} {\bibfnamefont {K.}~\bibnamefont
  {Hazzard}},\ }\href
  {http://iopscience.iop.org/article/10.1088/1367-2630/ad89f2} {\bibfield
  {journal} {\bibinfo  {journal} {New Journal of Physics}\ } (\bibinfo {year}
  {2024})}\BibitemShut {NoStop}%
\bibitem [{\citenamefont {Osterloh}\ \emph {et~al.}(2002)\citenamefont
  {Osterloh}, \citenamefont {Amico}, \citenamefont {Falci},\ and\ \citenamefont
  {Fazio}}]{Osterloh_2002}%
  \BibitemOpen
  \bibfield  {author} {\bibinfo {author} {\bibfnamefont {A.}~\bibnamefont
  {Osterloh}}, \bibinfo {author} {\bibfnamefont {L.}~\bibnamefont {Amico}},
  \bibinfo {author} {\bibfnamefont {G.}~\bibnamefont {Falci}}, \ and\ \bibinfo
  {author} {\bibfnamefont {R.}~\bibnamefont {Fazio}},\ }\href {\doibase
  10.1038/416608a} {\bibfield  {journal} {\bibinfo  {journal} {Nature}\
  }\textbf {\bibinfo {volume} {416}},\ \bibinfo {pages} {608} (\bibinfo {year}
  {2002})}\BibitemShut {NoStop}%
\bibitem [{\citenamefont {Osborne}\ and\ \citenamefont
  {Nielsen}(2002)}]{Osborne_2002}%
  \BibitemOpen
  \bibfield  {author} {\bibinfo {author} {\bibfnamefont {T.~J.}\ \bibnamefont
  {Osborne}}\ and\ \bibinfo {author} {\bibfnamefont {M.~A.}\ \bibnamefont
  {Nielsen}},\ }\href {\doibase 10.1103/PhysRevA.66.032110} {\bibfield
  {journal} {\bibinfo  {journal} {Phys. Rev. A}\ }\textbf {\bibinfo {volume}
  {66}},\ \bibinfo {pages} {032110} (\bibinfo {year} {2002})}\BibitemShut
  {NoStop}%
\bibitem [{\citenamefont {Vidal}\ \emph {et~al.}(2003)\citenamefont {Vidal},
  \citenamefont {Latorre}, \citenamefont {Rico},\ and\ \citenamefont
  {Kitaev}}]{Vidal_2003}%
  \BibitemOpen
  \bibfield  {author} {\bibinfo {author} {\bibfnamefont {G.}~\bibnamefont
  {Vidal}}, \bibinfo {author} {\bibfnamefont {J.~I.}\ \bibnamefont {Latorre}},
  \bibinfo {author} {\bibfnamefont {E.}~\bibnamefont {Rico}}, \ and\ \bibinfo
  {author} {\bibfnamefont {A.}~\bibnamefont {Kitaev}},\ }\href {\doibase
  10.1103/PhysRevLett.90.227902} {\bibfield  {journal} {\bibinfo  {journal}
  {Phys. Rev. Lett.}\ }\textbf {\bibinfo {volume} {90}},\ \bibinfo {pages}
  {227902} (\bibinfo {year} {2003})}\BibitemShut {NoStop}%
\bibitem [{\citenamefont {Filippone}\ \emph {et~al.}(2011)\citenamefont
  {Filippone}, \citenamefont {Dusuel},\ and\ \citenamefont
  {Vidal}}]{Filippone_2011}%
  \BibitemOpen
  \bibfield  {author} {\bibinfo {author} {\bibfnamefont {M.}~\bibnamefont
  {Filippone}}, \bibinfo {author} {\bibfnamefont {S.}~\bibnamefont {Dusuel}}, \
  and\ \bibinfo {author} {\bibfnamefont {J.}~\bibnamefont {Vidal}},\ }\href
  {\doibase 10.1103/PhysRevA.83.022327} {\bibfield  {journal} {\bibinfo
  {journal} {Phys. Rev. A}\ }\textbf {\bibinfo {volume} {83}},\ \bibinfo
  {pages} {022327} (\bibinfo {year} {2011})}\BibitemShut {NoStop}%
\bibitem [{\citenamefont {Botzung}\ and\ \citenamefont
  {Nataf}(2024{\natexlab{a}})}]{Botzung_2023_PRL}%
  \BibitemOpen
  \bibfield  {author} {\bibinfo {author} {\bibfnamefont {T.}~\bibnamefont
  {Botzung}}\ and\ \bibinfo {author} {\bibfnamefont {P.}~\bibnamefont
  {Nataf}},\ }\href {\doibase 10.1103/PhysRevLett.132.153001} {\bibfield
  {journal} {\bibinfo  {journal} {Phys. Rev. Lett.}\ }\textbf {\bibinfo
  {volume} {132}},\ \bibinfo {pages} {153001} (\bibinfo {year}
  {2024}{\natexlab{a}})}\BibitemShut {NoStop}%
\bibitem [{\citenamefont {Botzung}\ and\ \citenamefont
  {Nataf}(2024{\natexlab{b}})}]{Botzung_2023}%
  \BibitemOpen
  \bibfield  {author} {\bibinfo {author} {\bibfnamefont {T.}~\bibnamefont
  {Botzung}}\ and\ \bibinfo {author} {\bibfnamefont {P.}~\bibnamefont
  {Nataf}},\ }\href {\doibase 10.1103/PhysRevB.109.235131} {\bibfield
  {journal} {\bibinfo  {journal} {Phys. Rev. B}\ }\textbf {\bibinfo {volume}
  {109}},\ \bibinfo {pages} {235131} (\bibinfo {year}
  {2024}{\natexlab{b}})}\BibitemShut {NoStop}%
\bibitem [{\citenamefont {Itzykson}\ and\ \citenamefont
  {Nauenberg}(1966)}]{itzykson_unitarity_1966}%
  \BibitemOpen
  \bibfield  {author} {\bibinfo {author} {\bibfnamefont {C.}~\bibnamefont
  {Itzykson}}\ and\ \bibinfo {author} {\bibfnamefont {M.}~\bibnamefont
  {Nauenberg}},\ }\href {https://link.aps.org/doi/10.1103/RevModPhys.38.95}
  {\bibfield  {journal} {\bibinfo  {journal} {Rev. Mod. Phys.}\ }\textbf
  {\bibinfo {volume} {38}},\ \bibinfo {pages} {95} (\bibinfo {year}
  {1966})}\BibitemShut {NoStop}%
\bibitem [{\citenamefont {Weyl}(1925)}]{Weyl1925Dec}%
  \BibitemOpen
  \bibfield  {author} {\bibinfo {author} {\bibfnamefont {H.}~\bibnamefont
  {Weyl}},\ }\href {\doibase 10.1007/BF01506234} {\bibfield  {journal}
  {\bibinfo  {journal} {Math. Z.}\ }\textbf {\bibinfo {volume} {23}},\ \bibinfo
  {pages} {271} (\bibinfo {year} {1925})}\BibitemShut {NoStop}%
\bibitem [{\citenamefont {de~B.~Robinson}(1961)}]{Robinson1961}%
  \BibitemOpen
  \bibfield  {author} {\bibinfo {author} {\bibfnamefont {G.}~\bibnamefont
  {de~B.~Robinson}},\ }\href@noop {} {\emph {\bibinfo {title} {{Representation
  theory of the symmetric group}}}}\ (\bibinfo  {publisher} {Edinburgh: EUP},\
  \bibinfo {year} {1961})\BibitemShut {NoStop}%
\bibitem [{Note1()}]{Note1}%
  \BibitemOpen
  \bibinfo {note} {One can transform $H$ into $W^{\dagger }HW$, with
  $W=e^{i\protect \frac {\pi }{2}S_x}e^{-i\protect \frac {\pi
  }{2}S_y}$}\BibitemShut {NoStop}%
\bibitem [{\citenamefont {Pan}\ and\ \citenamefont {Draayer}(1999)}]{Pan_1999}%
  \BibitemOpen
  \bibfield  {author} {\bibinfo {author} {\bibfnamefont {F.}~\bibnamefont
  {Pan}}\ and\ \bibinfo {author} {\bibfnamefont {J.}~\bibnamefont {Draayer}},\
  }\href {\doibase https://doi.org/10.1016/S0370-2693(99)00191-4} {\bibfield
  {journal} {\bibinfo  {journal} {Physics Letters B}\ }\textbf {\bibinfo
  {volume} {451}},\ \bibinfo {pages} {1} (\bibinfo {year} {1999})}\BibitemShut
  {NoStop}%
\bibitem [{Note2()}]{Note2}%
  \BibitemOpen
  \bibinfo {note} {One has for instance $[H,W]=0$ with $W=e^{i\pi S_x}$ for $H$
  defined in Eq. \ref {Hamiltonian_LMG}.}\BibitemShut {Stop}%
\bibitem [{\citenamefont {Botet}\ \emph {et~al.}(1982)\citenamefont {Botet},
  \citenamefont {Jullien},\ and\ \citenamefont {Pfeuty}}]{Botet_1982}%
  \BibitemOpen
  \bibfield  {author} {\bibinfo {author} {\bibfnamefont {R.}~\bibnamefont
  {Botet}}, \bibinfo {author} {\bibfnamefont {R.}~\bibnamefont {Jullien}}, \
  and\ \bibinfo {author} {\bibfnamefont {P.}~\bibnamefont {Pfeuty}},\ }\href
  {\doibase 10.1103/PhysRevLett.49.478} {\bibfield  {journal} {\bibinfo
  {journal} {Phys. Rev. Lett.}\ }\textbf {\bibinfo {volume} {49}},\ \bibinfo
  {pages} {478} (\bibinfo {year} {1982})}\BibitemShut {NoStop}%
\bibitem [{\citenamefont {Botet}\ and\ \citenamefont
  {Jullien}(1983)}]{Botet_1983}%
  \BibitemOpen
  \bibfield  {author} {\bibinfo {author} {\bibfnamefont {R.}~\bibnamefont
  {Botet}}\ and\ \bibinfo {author} {\bibfnamefont {R.}~\bibnamefont
  {Jullien}},\ }\href {\doibase 10.1103/PhysRevB.28.3955} {\bibfield  {journal}
  {\bibinfo  {journal} {Phys. Rev. B}\ }\textbf {\bibinfo {volume} {28}},\
  \bibinfo {pages} {3955} (\bibinfo {year} {1983})}\BibitemShut {NoStop}%
\bibitem [{\citenamefont {Vidal}\ \emph
  {et~al.}(2004{\natexlab{a}})\citenamefont {Vidal}, \citenamefont {Palacios},\
  and\ \citenamefont {Mosseri}}]{Vidal_2004_second}%
  \BibitemOpen
  \bibfield  {author} {\bibinfo {author} {\bibfnamefont {J.}~\bibnamefont
  {Vidal}}, \bibinfo {author} {\bibfnamefont {G.}~\bibnamefont {Palacios}}, \
  and\ \bibinfo {author} {\bibfnamefont {R.}~\bibnamefont {Mosseri}},\ }\href
  {\doibase 10.1103/PhysRevA.69.022107} {\bibfield  {journal} {\bibinfo
  {journal} {Phys. Rev. A}\ }\textbf {\bibinfo {volume} {69}},\ \bibinfo
  {pages} {022107} (\bibinfo {year} {2004}{\natexlab{a}})}\BibitemShut
  {NoStop}%
\bibitem [{\citenamefont {Holstein}\ and\ \citenamefont
  {Primakoff}(1940)}]{Holstein_Primakoff}%
  \BibitemOpen
  \bibfield  {author} {\bibinfo {author} {\bibfnamefont {T.}~\bibnamefont
  {Holstein}}\ and\ \bibinfo {author} {\bibfnamefont {H.}~\bibnamefont
  {Primakoff}},\ }\href {\doibase 10.1103/PhysRev.58.1098} {\bibfield
  {journal} {\bibinfo  {journal} {Phys. Rev.}\ }\textbf {\bibinfo {volume}
  {58}},\ \bibinfo {pages} {1098} (\bibinfo {year} {1940})}\BibitemShut
  {NoStop}%
\bibitem [{\citenamefont {Dusuel}\ and\ \citenamefont
  {Vidal}(2004)}]{Vidal_PRL_2004}%
  \BibitemOpen
  \bibfield  {author} {\bibinfo {author} {\bibfnamefont {S.}~\bibnamefont
  {Dusuel}}\ and\ \bibinfo {author} {\bibfnamefont {J.}~\bibnamefont {Vidal}},\
  }\href {\doibase 10.1103/PhysRevLett.93.237204} {\bibfield  {journal}
  {\bibinfo  {journal} {Phys. Rev. Lett.}\ }\textbf {\bibinfo {volume} {93}},\
  \bibinfo {pages} {237204} (\bibinfo {year} {2004})}\BibitemShut {NoStop}%
\bibitem [{\citenamefont {Rosensteel}\ \emph {et~al.}(2007)\citenamefont
  {Rosensteel}, \citenamefont {Rowe},\ and\ \citenamefont
  {Ho}}]{Rosensteel_2008}%
  \BibitemOpen
  \bibfield  {author} {\bibinfo {author} {\bibfnamefont {G.}~\bibnamefont
  {Rosensteel}}, \bibinfo {author} {\bibfnamefont {D.~J.}\ \bibnamefont
  {Rowe}}, \ and\ \bibinfo {author} {\bibfnamefont {S.~Y.}\ \bibnamefont
  {Ho}},\ }\href {\doibase 10.1088/1751-8113/41/2/025208} {\bibfield  {journal}
  {\bibinfo  {journal} {Journal of Physics A: Mathematical and Theoretical}\
  }\textbf {\bibinfo {volume} {41}},\ \bibinfo {pages} {025208} (\bibinfo
  {year} {2007})}\BibitemShut {NoStop}%
\bibitem [{\citenamefont {Newman}\ and\ \citenamefont
  {Schulman}(1977)}]{Newman_1977}%
  \BibitemOpen
  \bibfield  {author} {\bibinfo {author} {\bibfnamefont {C.~M.}\ \bibnamefont
  {Newman}}\ and\ \bibinfo {author} {\bibfnamefont {L.~S.}\ \bibnamefont
  {Schulman}},\ }\href {\doibase 10.1063/1.523131} {\bibfield  {journal}
  {\bibinfo  {journal} {Journal of Mathematical Physics}\ }\textbf {\bibinfo
  {volume} {18}},\ \bibinfo {pages} {23} (\bibinfo {year} {1977})},\ \Eprint
  {http://arxiv.org/abs/https://pubs.aip.org/aip/jmp/article-pdf/18/1/23/19251057/23\_1\_online.pdf}
  {https://pubs.aip.org/aip/jmp/article-pdf/18/1/23/19251057/23\_1\_online.pdf}
  \BibitemShut {NoStop}%
\bibitem [{\citenamefont {Vidal}\ \emph
  {et~al.}(2004{\natexlab{b}})\citenamefont {Vidal}, \citenamefont {Mosseri},\
  and\ \citenamefont {Dukelsky}}]{Vidal_2004_first}%
  \BibitemOpen
  \bibfield  {author} {\bibinfo {author} {\bibfnamefont {J.}~\bibnamefont
  {Vidal}}, \bibinfo {author} {\bibfnamefont {R.}~\bibnamefont {Mosseri}}, \
  and\ \bibinfo {author} {\bibfnamefont {J.}~\bibnamefont {Dukelsky}},\ }\href
  {\doibase 10.1103/PhysRevA.69.054101} {\bibfield  {journal} {\bibinfo
  {journal} {Phys. Rev. A}\ }\textbf {\bibinfo {volume} {69}},\ \bibinfo
  {pages} {054101} (\bibinfo {year} {2004}{\natexlab{b}})}\BibitemShut
  {NoStop}%
\bibitem [{\citenamefont {Vidal}\ \emph {et~al.}(2007)\citenamefont {Vidal},
  \citenamefont {Dusuel},\ and\ \citenamefont {Barthel}}]{Vidal_2007}%
  \BibitemOpen
  \bibfield  {author} {\bibinfo {author} {\bibfnamefont {J.}~\bibnamefont
  {Vidal}}, \bibinfo {author} {\bibfnamefont {S.}~\bibnamefont {Dusuel}}, \
  and\ \bibinfo {author} {\bibfnamefont {T.}~\bibnamefont {Barthel}},\ }\href
  {\doibase 10.1088/1742-5468/2007/01/P01015} {\bibfield  {journal} {\bibinfo
  {journal} {Journal of Statistical Mechanics: Theory and Experiment}\ }\textbf
  {\bibinfo {volume} {2007}},\ \bibinfo {pages} {P01015} (\bibinfo {year}
  {2007})}\BibitemShut {NoStop}%
\bibitem [{\citenamefont {Hines}\ \emph {et~al.}(2003)\citenamefont {Hines},
  \citenamefont {McKenzie},\ and\ \citenamefont {Milburn}}]{Hines_2003}%
  \BibitemOpen
  \bibfield  {author} {\bibinfo {author} {\bibfnamefont {A.~P.}\ \bibnamefont
  {Hines}}, \bibinfo {author} {\bibfnamefont {R.~H.}\ \bibnamefont {McKenzie}},
  \ and\ \bibinfo {author} {\bibfnamefont {G.~J.}\ \bibnamefont {Milburn}},\
  }\href {\doibase 10.1103/PhysRevA.67.013609} {\bibfield  {journal} {\bibinfo
  {journal} {Phys. Rev. A}\ }\textbf {\bibinfo {volume} {67}},\ \bibinfo
  {pages} {013609} (\bibinfo {year} {2003})}\BibitemShut {NoStop}%
\bibitem [{\citenamefont {Fuhringer}\ \emph {et~al.}(2008)\citenamefont
  {Fuhringer}, \citenamefont {Rachel}, \citenamefont {Thomale}, \citenamefont
  {Greiter},\ and\ \citenamefont {Schmitteckert}}]{fuhringer2008}%
  \BibitemOpen
  \bibfield  {author} {\bibinfo {author} {\bibfnamefont {M.}~\bibnamefont
  {Fuhringer}}, \bibinfo {author} {\bibfnamefont {S.}~\bibnamefont {Rachel}},
  \bibinfo {author} {\bibfnamefont {R.}~\bibnamefont {Thomale}}, \bibinfo
  {author} {\bibfnamefont {M.}~\bibnamefont {Greiter}}, \ and\ \bibinfo
  {author} {\bibfnamefont {P.}~\bibnamefont {Schmitteckert}},\ }\href {\doibase
  10.1002/andp.200810326} {\bibfield  {journal} {\bibinfo  {journal} {Annalen
  der Physik}\ }\textbf {\bibinfo {volume} {17}},\ \bibinfo {pages} {922}
  (\bibinfo {year} {2008})}\BibitemShut {NoStop}%
\bibitem [{\citenamefont {Szirmai}\ \emph {et~al.}(2008)\citenamefont
  {Szirmai}, \citenamefont {Legeza},\ and\ \citenamefont
  {S\'olyom}}]{Szirmai_2008}%
  \BibitemOpen
  \bibfield  {author} {\bibinfo {author} {\bibfnamefont {E.}~\bibnamefont
  {Szirmai}}, \bibinfo {author} {\bibfnamefont {O.}~\bibnamefont {Legeza}}, \
  and\ \bibinfo {author} {\bibfnamefont {J.}~\bibnamefont {S\'olyom}},\ }\href
  {\doibase 10.1103/PhysRevB.77.045106} {\bibfield  {journal} {\bibinfo
  {journal} {Phys. Rev. B}\ }\textbf {\bibinfo {volume} {77}},\ \bibinfo
  {pages} {045106} (\bibinfo {year} {2008})}\BibitemShut {NoStop}%
\bibitem [{\citenamefont {D'Emidio}\ \emph {et~al.}(2015)\citenamefont
  {D'Emidio}, \citenamefont {Block},\ and\ \citenamefont
  {Kaul}}]{Demidio_2015}%
  \BibitemOpen
  \bibfield  {author} {\bibinfo {author} {\bibfnamefont {J.}~\bibnamefont
  {D'Emidio}}, \bibinfo {author} {\bibfnamefont {M.~S.}\ \bibnamefont {Block}},
  \ and\ \bibinfo {author} {\bibfnamefont {R.~K.}\ \bibnamefont {Kaul}},\
  }\href {\doibase 10.1103/PhysRevB.92.054411} {\bibfield  {journal} {\bibinfo
  {journal} {Phys. Rev. B}\ }\textbf {\bibinfo {volume} {92}},\ \bibinfo
  {pages} {054411} (\bibinfo {year} {2015})}\BibitemShut {NoStop}%
\bibitem [{\citenamefont {Nataf}\ and\ \citenamefont
  {Mila}(2018)}]{nataf_density_2018}%
  \BibitemOpen
  \bibfield  {author} {\bibinfo {author} {\bibfnamefont {P.}~\bibnamefont
  {Nataf}}\ and\ \bibinfo {author} {\bibfnamefont {F.}~\bibnamefont {Mila}},\
  }\href {https://link.aps.org/doi/10.1103/PhysRevB.97.134420} {\bibfield
  {journal} {\bibinfo  {journal} {Phys. Rev. B}\ }\textbf {\bibinfo {volume}
  {97}},\ \bibinfo {pages} {134420} (\bibinfo {year} {2018})}\BibitemShut
  {NoStop}%
\bibitem [{\citenamefont {Shi}\ \emph {et~al.}(2024)\citenamefont {Shi},
  \citenamefont {Dai}, \citenamefont {Zhou},\ and\ \citenamefont
  {McCulloch}}]{shi2024}%
  \BibitemOpen
  \bibfield  {author} {\bibinfo {author} {\bibfnamefont {Q.-Q.}\ \bibnamefont
  {Shi}}, \bibinfo {author} {\bibfnamefont {Y.-W.}\ \bibnamefont {Dai}},
  \bibinfo {author} {\bibfnamefont {H.-Q.}\ \bibnamefont {Zhou}}, \ and\
  \bibinfo {author} {\bibfnamefont {I.}~\bibnamefont {McCulloch}},\ }\href
  {https://arxiv.org/abs/2201.01071} {\bibfield  {journal} {\bibinfo  {journal}
  {arXiv/Condmat}\ } (\bibinfo {year} {2024})},\ \Eprint
  {http://arxiv.org/abs/2201.01071} {arXiv:2201.01071 [cond-mat.str-el]}
  \BibitemShut {NoStop}%
\bibitem [{\citenamefont {Calabrese}\ and\ \citenamefont
  {Cardy}(2004)}]{calabrese_entanglement_2004}%
  \BibitemOpen
  \bibfield  {author} {\bibinfo {author} {\bibfnamefont {P.}~\bibnamefont
  {Calabrese}}\ and\ \bibinfo {author} {\bibfnamefont {J.}~\bibnamefont
  {Cardy}},\ }\href
  {https://iopscience.iop.org/article/10.1088/1742-5468/2004/06/P06002/meta}
  {\bibfield  {journal} {\bibinfo  {journal} {Journal of Statistical Mechanics:
  Theory and Experiment}\ }\textbf {\bibinfo {volume} {2004}},\ \bibinfo
  {pages} {P06002} (\bibinfo {year} {2004})}\BibitemShut {NoStop}%
\bibitem [{\citenamefont {Huckans}\ \emph {et~al.}(2009)\citenamefont
  {Huckans}, \citenamefont {Williams}, \citenamefont {Hazlett}, \citenamefont
  {Stites},\ and\ \citenamefont {O'Hara}}]{Huckans_2009}%
  \BibitemOpen
  \bibfield  {author} {\bibinfo {author} {\bibfnamefont {J.~H.}\ \bibnamefont
  {Huckans}}, \bibinfo {author} {\bibfnamefont {J.~R.}\ \bibnamefont
  {Williams}}, \bibinfo {author} {\bibfnamefont {E.~L.}\ \bibnamefont
  {Hazlett}}, \bibinfo {author} {\bibfnamefont {R.~W.}\ \bibnamefont {Stites}},
  \ and\ \bibinfo {author} {\bibfnamefont {K.~M.}\ \bibnamefont {O'Hara}},\
  }\href {\doibase 10.1103/PhysRevLett.102.165302} {\bibfield  {journal}
  {\bibinfo  {journal} {Phys. Rev. Lett.}\ }\textbf {\bibinfo {volume} {102}},\
  \bibinfo {pages} {165302} (\bibinfo {year} {2009})}\BibitemShut {NoStop}%
\bibitem [{\citenamefont {Pagano}\ \emph {et~al.}(2015)\citenamefont {Pagano},
  \citenamefont {Mancini}, \citenamefont {Cappellini}, \citenamefont {Livi},
  \citenamefont {Sias}, \citenamefont {Catani}, \citenamefont {Inguscio},\ and\
  \citenamefont {Fallani}}]{Pagano_2015}%
  \BibitemOpen
  \bibfield  {author} {\bibinfo {author} {\bibfnamefont {G.}~\bibnamefont
  {Pagano}}, \bibinfo {author} {\bibfnamefont {M.}~\bibnamefont {Mancini}},
  \bibinfo {author} {\bibfnamefont {G.}~\bibnamefont {Cappellini}}, \bibinfo
  {author} {\bibfnamefont {L.}~\bibnamefont {Livi}}, \bibinfo {author}
  {\bibfnamefont {C.}~\bibnamefont {Sias}}, \bibinfo {author} {\bibfnamefont
  {J.}~\bibnamefont {Catani}}, \bibinfo {author} {\bibfnamefont
  {M.}~\bibnamefont {Inguscio}}, \ and\ \bibinfo {author} {\bibfnamefont
  {L.}~\bibnamefont {Fallani}},\ }\href {\doibase
  10.1103/PhysRevLett.115.265301} {\bibfield  {journal} {\bibinfo  {journal}
  {Phys. Rev. Lett.}\ }\textbf {\bibinfo {volume} {115}},\ \bibinfo {pages}
  {265301} (\bibinfo {year} {2015})}\BibitemShut {NoStop}%
\bibitem [{\citenamefont {H\"ofer}\ \emph {et~al.}(2015)\citenamefont
  {H\"ofer}, \citenamefont {Riegger}, \citenamefont {Scazza}, \citenamefont
  {Hofrichter}, \citenamefont {Fernandes}, \citenamefont {Parish},
  \citenamefont {Levinsen}, \citenamefont {Bloch},\ and\ \citenamefont
  {F\"olling}}]{Hofer_2015}%
  \BibitemOpen
  \bibfield  {author} {\bibinfo {author} {\bibfnamefont {M.}~\bibnamefont
  {H\"ofer}}, \bibinfo {author} {\bibfnamefont {L.}~\bibnamefont {Riegger}},
  \bibinfo {author} {\bibfnamefont {F.}~\bibnamefont {Scazza}}, \bibinfo
  {author} {\bibfnamefont {C.}~\bibnamefont {Hofrichter}}, \bibinfo {author}
  {\bibfnamefont {D.~R.}\ \bibnamefont {Fernandes}}, \bibinfo {author}
  {\bibfnamefont {M.~M.}\ \bibnamefont {Parish}}, \bibinfo {author}
  {\bibfnamefont {J.}~\bibnamefont {Levinsen}}, \bibinfo {author}
  {\bibfnamefont {I.}~\bibnamefont {Bloch}}, \ and\ \bibinfo {author}
  {\bibfnamefont {S.}~\bibnamefont {F\"olling}},\ }\href {\doibase
  10.1103/PhysRevLett.115.265302} {\bibfield  {journal} {\bibinfo  {journal}
  {Phys. Rev. Lett.}\ }\textbf {\bibinfo {volume} {115}},\ \bibinfo {pages}
  {265302} (\bibinfo {year} {2015})}\BibitemShut {NoStop}%
\bibitem [{\citenamefont {Mukherjee}\ and\ \citenamefont
  {Hutson}(2024)}]{Mukherjee_2024_arxiv}%
  \BibitemOpen
  \bibfield  {author} {\bibinfo {author} {\bibfnamefont {B.}~\bibnamefont
  {Mukherjee}}\ and\ \bibinfo {author} {\bibfnamefont {J.~M.}\ \bibnamefont
  {Hutson}},\ }\href {https://arxiv.org/abs/2410.19068} {\bibfield  {journal}
  {\bibinfo  {journal} {arXiv/Condmat}\ } (\bibinfo {year} {2024})},\ \Eprint
  {http://arxiv.org/abs/2410.19068} {arXiv:2410.19068 [quant-ph]} \BibitemShut
  {NoStop}%
\bibitem [{\citenamefont {Unanyan}\ and\ \citenamefont
  {Fleischhauer}(2003)}]{Unanyan_2003}%
  \BibitemOpen
  \bibfield  {author} {\bibinfo {author} {\bibfnamefont {R.~G.}\ \bibnamefont
  {Unanyan}}\ and\ \bibinfo {author} {\bibfnamefont {M.}~\bibnamefont
  {Fleischhauer}},\ }\href {\doibase 10.1103/PhysRevLett.90.133601} {\bibfield
  {journal} {\bibinfo  {journal} {Phys. Rev. Lett.}\ }\textbf {\bibinfo
  {volume} {90}},\ \bibinfo {pages} {133601} (\bibinfo {year}
  {2003})}\BibitemShut {NoStop}%
\end{thebibliography}%

\end{document}